\documentclass{emulateapj}
\slugcomment{{\sc Accepted to ApJ:} March 22, 2012}

\usepackage{multirow}

\newcommand{\hmpc}{h^{-1}\, {\rm{Mpc}}}
\newcommand{\msol}{{\rm M}_{\odot}} 
\newcommand{\rc}{r_{500}} 
\newcommand{\trc}{r_{200}} 
\newcommand{\tfrc}{r_{2500}}
\newcommand{\Gyr}{\,\mathrm{Gyr}}

\begin{document}
\title{Evolution of the Merger Induced Hydrostatic Mass Bias in Galaxy Clusters}

\author{Kaylea Nelson\altaffilmark{1}, Douglas H. Rudd\altaffilmark{2,3}, Laurie Shaw\altaffilmark{2,3}, Daisuke Nagai\altaffilmark{1,2,3}}

\altaffiltext{1}{Department of Astronomy, Yale University, New Haven, CT, 06520}
\altaffiltext{2}{Department of Physics, Yale University, New Haven, CT 06520}
\altaffiltext{3}{Yale Center for Astronomy \& Astrophysics, Yale University, New Haven, CT 06520}
\email{kaylea.nelson@yale.edu}

\begin{abstract}
In this work, we examine the effects of mergers on the hydrostatic mass estimate of galaxy clusters using high-resolution Eulerian cosmological simulations.  We utilize merger trees to isolate the last merger for each cluster in our sample and follow the time evolution of the hydrostatic mass bias as the systems relax.  We find that during a merger, a shock propagates outward from the parent cluster, resulting in an overestimate in the hydrostatic mass bias.  After the merger, as a cluster relaxes, the bias in hydrostatic mass estimate decreases but remains at a level of -5--10\% with 15--20\% scatter within $\rc$. We also investigate the post-merger evolution of the pressure support from bulk motions, a dominant cause of this residual mass bias.  At $\rc$, the contribution from random motions peaks at 30\% of the total pressure during the merger and quickly decays to $\sim$10--15\% as a cluster relaxes.  Additionally, we use a measure of the random motion pressure to correct the hydrostatic mass estimate.  We discover that 4 Gyr after mergers, the direct effects of the merger event on the hydrostatic mass bias have become negligible. Thereafter, the mass bias is primarily due to residual bulk motions in the gas which are not accounted for in the hydrostatic equilibrium equation. We present a hydrostatic mass bias correction method that can recover the unbiased cluster mass for relaxed clusters with 9\% scatter at $\rc$ and 11\% scatter in the outskirts, within $\trc$.

\end{abstract}
  
\section{Introduction}  

Over the last decade, measurements of the abundance of galaxy clusters
have begun to provide important constraints on cosmological parameters
\citep{Vikhlinin:2009, Mantz:2010, Rozo:2010, Vanderlinde:2010, Sehgal:2010}. Clusters 
trace the highest peaks in the matter density field; their abundance as 
a function of mass and redshift is therefore extremely sensitive to the 
growth rate of density perturbations as well as the expansion rate of the 
Universe.

The power of cluster number counts as a cosmological probe is principally 
limited by the accuracy with which their total mass can be measured. 
Cluster mass can be estimated using a wide variety of techniques, from 
measurements of the number or velocity dispersion of the cluster member galaxies, 
strong and weak gravitational lensing of background sources, and 
measurements of their X-ray emission or Sunyaev-Zel'dovich (SZ) signal. 
The two latter methods utilize the temperature and density of the hot 
plasma, or intra-cluster medium (ICM), that pervades cluster environments 
and dominates their baryonic mass. The thermal properties of the ICM are 
very sensitive to the depth of the gravitational potential of the cluster 
host dark matter halo, and therefore to the total cluster mass. In 
particular, by measuring the radial density and temperature profiles of 
the ICM from X-ray and SZ observations, the equation of hydrostatic 
equilibrium (HSE) can be solved to obtain the total matter profile.

Application of the equation of HSE is based on the assumption that
the ICM is both spherically symmetric and in hydrostatic equilibrium.
However, clusters are dynamically young systems and obtain a significant
fraction of their mass through mergers.  Therefore the accuracy of 
this assumption must be questioned. Recent cosmological simulations have 
suggested that even in relaxed clusters the HSE method may systematically 
underestimate the total cluster mass by $\sim$10--20\% 
\citep{Kay:2004, Rasia:2006, Nagai:2007a}. It has been demonstrated that 
this bias is a consequence of neglecting the contribution of residual bulk 
motions in the ICM to the dynamical stability of relaxed clusters 
\citep{Kay:2004, Rasia:2004, Jeltema:2008, Piffaretti:2008, Lau:2009}. Typically, the HSE technique 
only accounts for support from thermal pressure; the non-thermal
contribution from bulk motions and turbulence is difficult to measure 
directly. The evolution of these bulk motions, their generation through 
mergers, and the potential effect on the corresponding evolution of the 
mass bias are all very poorly understood.

Several recent works use cosmological simulations of clusters 
to investigate the effects of dynamical state on the
deviation from hydrostatic equilibrium by examining a number of
clusters at single redshifts.  \citet{Nagai:2007a} demonstrate that
the mass bias increases rapidly with radius and is 20\% or greater 
at all radii in unrelaxed systems. Utilizing a 
large sample of over 100 clusters, \citet{Piffaretti:2008} show that 
the hydrostatic mass bias is larger for the less relaxed, recently
merged clusters, and also that the scatter in the bias increases 
with the degree of disturbance in the systems.  

In an attempt to incorporate observational uncertainties, 
\citet{Meneghetti:2010} investigate the effect of line 
of sight projections using mock lensing and mock X-ray observations and were
able to recover the bias measured directly in simulations. In an extension of \citet{Meneghetti:2010},
\citet{Rasia:2012} also investigate the effects of cluster morphology, environment, temperature inhomogeneity,
and mass on the mass biases. Moreover, both works
show agreement with an observational comparison of the X-ray hydrostatic mass 
estimate and weak lensing mass estimates by \citet{Mahdavi:2008}, namely 
that the HSE mass derived from {\it Chandra} X-ray measurements is biased
low by $\sim20\%$ at $\rc$\footnotemark\  compared to weak gravitational lensing measurements.
Similarly, \citet{Zhang:2010} investigate the effect of dynamical state on the 
bias between the mass measurement methods and find a $\sim10\%$ underestimate 
in the HSE mass at $\rc$ in undisturbed clusters but also find that in disturbed clusters 
the mass is overestimated by $\sim6\%$ at the same radius.

\footnotetext[4]{Throughout this paper, we refer to total mass and ICM properties within
radii which correspond to fixed overdensities $\Delta$ relative to the
critical density at that redshift, such that 
$M_{\Delta} (r_{\Delta}) = \Delta(4/3)\pi r^3_{\Delta}\rho_c(z)$.} 

In this paper, we use high-resolution cosmological simulations to
investigate the time evolution of the hydrostatic mass bias and
non-thermal pressure support in the ICM as a function of time since
the latest merger.  While a number of recent studies have explored the
effects of mergers and dynamical state on the hydrostatic mass
estimate \citep[e.g.][]{ Kay:2004, Rasia:2006, Nagai:2007a,
  Lau:2009,Poole:2006, Piffaretti:2008}, most only explored these
effects at a single epoch or in idealized mergers. Our simulations and
merger trees enable us to characterize the dynamical state of clusters
robustly and quantitatively, and more effectively follow the evolution
of the hydrostatic mass estimate during and after mergers in the
context of realistic cosmological simulations. 
We investigate the source of the hydrostatic mass bias by
following the post-merger evolution of the non-thermal pressure
component of the ICM contributed primarily by random gas motions\footnotemark . 
We find that during mergers, 
the bulk motions supply about 30\% of the pressure support in clusters
within $\rc$.  While the random motion pressure quickly
decays, we also find that after 9 Gyr, these gas motions still
provide about 10\% of the pressure support.  By taking the random motion
pressure into account in a correction factor, we are able to
recover the true mass of relaxed clusters, clusters which have not
undergone a merger in more than 4 Gyr, to within a few percent. We
conclude by presenting a potential method for estimating this
correction factor using the thermal pressure profile.

\footnotetext[5]{Additional contributions from cosmic rays and magnetic field 
are not included.}

In Section~\ref{sec:simulations}, we describe our cosmological cluster
simulations. In Section~\ref{sec:individual}, we present our analysis of 
the time evolution of the hydrostatic mass bias.  In Section~\ref{sec:pturb}
we explore the connection between non-thermal pressure and present our model
for correcting the mass bias.  In Section~\ref{sec:conclusions}, we summarize 
and discuss our findings.

\section{Simulations}
\label{sec:simulations}

We analyze a sample of 16 simulated groups and clusters of galaxies presented 
previously in \citet{Nagai:2007a} and \citet{Nagai:2007b}, and we refer 
the reader to these papers for more details. We briefly summarize the relevant 
parameters of the simulations here.

The simulations are performed using the Adaptive Refinement Tree (ART) $N$-body$+$gas-dynamics code 
\citep{Kravtsov:1999,Kravtsov:2002,Rudd:2008}, which is an Eulerian code that uses adaptive refinement in space 
and time, and non-adaptive refinement in mass \citep{Klypin:2001} to achieve the dynamic range necessary to 
resolve the cores of halos formed in self-consistent cosmological simulations. The simulations assume 
a flat {$\Lambda$}CDM model: $\Omega_{\rm m}=1-\Omega_{\Lambda}=0.3$, $\Omega_{\rm b}=0.04286$, $h=0.7$ and $\sigma_8=0.9$, 
where the Hubble constant is defined as $100h{\ \rm km\ s^{-1}\ Mpc^{-1}}$, and $\sigma_8$ is the mass 
variance within spheres of radius $8\,h^{-1}$~Mpc. The box size for 
CL101--CL107 is $120\,h^{-1}$~Mpc comoving on a side and $80\,h^{-1}$~Mpc comoving for 
CL3--CL24. All simulations were run using a uniform $128^3$ grid with 8 levels of mesh 
refinement, corresponding to peak spatial resolution of $\sim 7\,h^{-1}$~kpc and $5\,h^{-1}$~kpc 
for the two box sizes respectively.  Only the regions $\sim 3-10\,h^{-1}$~Mpc surrounding each cluster 
center were adaptively refined. The dark matter (DM) particle mass in the region around each cluster was 
$m_{\rm p} \approxeq 9.1\times 10^{8}\,h^{-1}\, {\rm M_{\odot}}$ for CL101--CL107 and
$ m_{\rm p} \approxeq 2.7\times 10^{8}\,h^{-1}\,{\rm M_{\odot}}$ for CL3--CL24, while other regions 
were simulated with a lower mass resolution.  In order to isolate the effect of mergers
on the ICM we use simulations that treat only gravity and non-radiative hydrodynamics. 
\citet{Lau:2009} show that the exclusion of cooling and star formation does not have a 
strong effect on the total contribution of gas motions to the HSE reconstruction of mass
profiles outside of cluster cores. 

%*********************************************************************
\begin{deluxetable}{lcccc}
\tablecolumns{5}
\tablecaption{List of Simulated Clusters\label{table:clusters}}
\tablehead{
\multirow{2}{*}{Name\tablenotemark{1}\hspace{0.05\linewidth}} &
\multicolumn{1}{c}{$r_{500}$} &
\multicolumn{1}{c}{$M_{500}$} &
\multicolumn{1}{c}{$t_{\mathrm{merger}}$\tablenotemark{2}} &
\multicolumn{1}{c}{Merger Ratio}
\\
& $(\hmpc)$ & $(10^{14} h^{-1} \msol)$ & $\Gyr$ & }
\startdata
CL101 & 1.128 & 8.31 &  2.19 & 0.49  \\
CL102 & 0.967 & 5.24 &  3.09 & 0.27  \\
CL103 & 0.975 & 5.37 &  2.00 & 0.81  \\
CL104 & 0.970 & 5.29 &  8.72 & 0.24  \\
CL105 & 0.902 & 4.26 & 8.53 & 0.21  \\
CL106 & 0.854 & 3.61 & 1.64 & 0.31  \\
CL107 & 0.766 & 2.60 &  5.98\tablenotemark{3} & 0.60  \\
CL3 & 0.721 & 2.17 &  6.14 & 0.18    \\
CL5 & 0.546 & 0.94 &  5.88 & 0.33    \\
CL6 & 0.590 & 1.19 &  5.86 & 0.71    \\
CL7 & 0.589 & 1.19 &  11.28 & 0.17    \\
CL9 & 0.469 & 0.60 & 6.90 & 0.75    \\
CL10 & 0.475 & 0.62 & 7.18  & 0.50  \\
CL11 & 0.425 & 0.45 & 9.62  & 0.80  \\
CL14 & 0.483 & 0.66 &  4.64 & 0.25   \\
CL24 & 0.394 & 0.35 & 1.95  & 0.84  
\enddata
\tablenotetext{1}{Cluster labels correspond to those used in \citet{Nagai:2007a}.}
\tablenotetext{2}{Time elapsed at $z = 0$ since last major merger.}
\tablenotetext{3}{At $z$ = 0, CL107 is in the prelimary phases of a major merger with a halo that has not yet entered $\trc$ and therefore is classified as an unrelaxed cluster despite having $t_{merger}$ = 5.98 Gyr.}
\end{deluxetable}
%*********************************************************************

Each cluster has 30 complete outputs in the range $z \le 4$, 
with the majority occurring equally spaced in expansion factor after $z = 1$ with 
$\Delta a = \Delta(1+z)^{-1} = 0.025$. The dark matter particles, used to identify cluster mergers, are 
output at a much finer spacing with $\approx$300 outputs over the cluster history.  This 
allows for a robust identification of the cluster progenitor and merger timing.

\subsection{Halo Finder}

A variant of the bound density maxima halo finding algorithm
\citep{Klypin:1999} is used to identify halos and the subhalos within
them. The details of the algorithm and parameters used in the halo
finder can be found in \citet{Kravtsov:2004a}. The main steps of the
algorithm are identification of local density peaks (potential halo
centers) and analysis of the density distribution and velocities of
the surrounding particles to test whether a given peak corresponds to
a gravitationally bound clump. More specifically, we construct
density, circular velocity, and velocity dispersion profiles around
each center and iteratively remove unbound particles using the
procedure outlined in \citet{Klypin:1999}. 

\subsection{Merger Trees}

To examine the evolution of cluster properties and the hydrostatic
mass bias, we track the evolution of each cluster's most massive
progenitor. We identify the main progenitor of each $z = 0$ cluster by
iteratively following the 10\% most bound dark matter particles at
each epoch.  Mergers are located by identifying objects that cross
$r_{200}$ of the parent cluster.  We determine the mass ratio of the merger at an epoch when the 
two systems are sufficiently separated to avoid impacting their respective
mass profiles, herein we use the epoch at which they initially overlap at a radius
of $r_{200}$. We use the ratio of $M_{500}$ at this epoch to identify major mergers, 
herein defined as a mass ratio greater than 1:6.  Merging events with the mass ratios 
smaller than this are defined as minor mergers.
While this ratio is somewhat lower than used in other studies \citep[e.g.][]{Gottlober:2001}, 
careful exploration of the merger ratio has shown that this threshold best separates mergers 
that have a strong impact on the measured ICM properties from those that do
not.  We examine the evolution of these properties as a function of
time since the last major merger, $t_{\mathrm{merger}}$, defined as
the epoch at which the center of the merging cluster first crosses $r_{200}$ of the
parent cluster.

\subsection{Cluster Profiles}
\label{sec:clusterprofiles}

For each simulation output we measure properties of the ICM under the
assumption of spherical symmetry following the methods described in
\cite{Lau:2009}.  A brief description of the methods used is included
here.  We focus in this work on the estimate of cluster mass provided
by the thermal hydrostatic equilibrium equation,
\begin{equation}
\label{eq:hse}
    M_{\mathrm{HSE}} (< r) = - \frac{r^2}{G\rho_{gas}} \frac{dP_{th}}{dr},
\end{equation}
\noindent where $M_{\mathrm{HSE}} (<r)$ is the estimated mass within
radius, $r$, and $\rho_{gas}(r)$ and $P_{th}(r)$ are the gas density
and thermal pressure, respectively. These are measured within
spherical shells, logarithmically-spaced in radius from the cluster
center.

We also measure the contribution of isotropic pressure from random gas motions,
\begin{equation}
P_{\mathrm{rand}} = \frac{1}{3} \rho_{\mathrm{gas}} \left( \sigma_r^2 + \sigma_t^2 \right),
\end{equation}
where $\sigma_r^2$ and $\sigma_t^2$ are radial and tangential velocity dispersions
measured in spherical shells, using the peculiar velocity of the cluster dark matter
within $\rc$ to define a rest frame.  In the absence of radiative cooling, rotational
velocities are negligible outside $\sim0.2\rc$, as was seen in \cite{Lau:2009}, however
for completeness we treat their contribution to the effective pressure support separately.

Following \citet{Lau:2009}, we treat the random gas motions using approach similar to 
the treatment of the dynamical collisionless system using the Jeans equation 
\citep{Binney:2008}.  The total mass of the cluster within a radius $r$ can be split into 
separate components corresponding to different terms of the Jeans equation, 
\begin{equation}
\label{eq:MJeans}
M_{\mathrm{tot}}(<r) = M_{\mathrm{HSE}} + M_{\mathrm{rand}} + M_{\mathrm{rot}}.
\end{equation}
The streaming and cross terms are small, and we neglect them throughout this work. 
The rotational component, which contributes significantly to the pressure support in 
the core ($<$ 0.5 $\rc$), is given by
\begin{equation}
M_{\mathrm{rot}}(<r) = \frac{r \bar{v}_t^2}{G},
\end{equation}
where $\bar{v}_t$ is the mean tangential velocity. The component from random gas motions is given by
\begin{equation}
M_{\mathrm{rand}}(<r) = \frac{-r^2}{G\rho_{\mathrm{gas}}} \left( \frac{\partial\left(\rho_{\mathrm{gas}} \sigma_r^2 \right)}{\partial r}\right) - \frac{r}{G}\left(2\sigma_r^2 - \sigma_t^2\right).
\end{equation}

\noindent Note that spherical symmetry for the gravitational potential and steady state are assumed in deriving the expression for mass from the Jeans equation, and all the physical quantities at a given radius are averages over a radial shell.

To avoid biases due to the presence of substructure and to maintain consistency with 
\cite{Lau:2009}, in measuring $M_{\mathrm{\mathrm{HSE}}}$ and $P_{\mathrm{rand}}$ 
we remove gas from within the tidal radius of subhalos of mass greater than 
$10^{12} h^{-1} \msol$. We further reduce small scale fluctuations in pressure and 
pressure gradient profiles by smoothing with a Savitzky-Golay filter. We have explicitly 
checked that neither subhalo removal nor smoothing procedures have a significant 
effect on our results.

It is important to note that all radial quantities in this paper,
including $M_{\mathrm{HSE}}$, are measured within the corresponding
``true" radius, $r_{\mathrm{true}}$, as measured directly from the
simulations. \citet{Lau:2009} shows that if $M_{\mathrm{HSE}}$ is
instead measured within the radius as estimated from the HSE equation,
$r_{\mathrm{est}}$, the hydrostatic mass bias within $\rc$ is increased by 
$\sim$4\% for relaxed systems \citep[see also][for results based on mock 
X-ray maps]{Nagai:2007a}.

\section{Evolution of the Hydrostatic Mass Bias}
\label{sec:individual}

\begin{figure*}[t]
\epsscale{1.1}

\centerline{\plotone{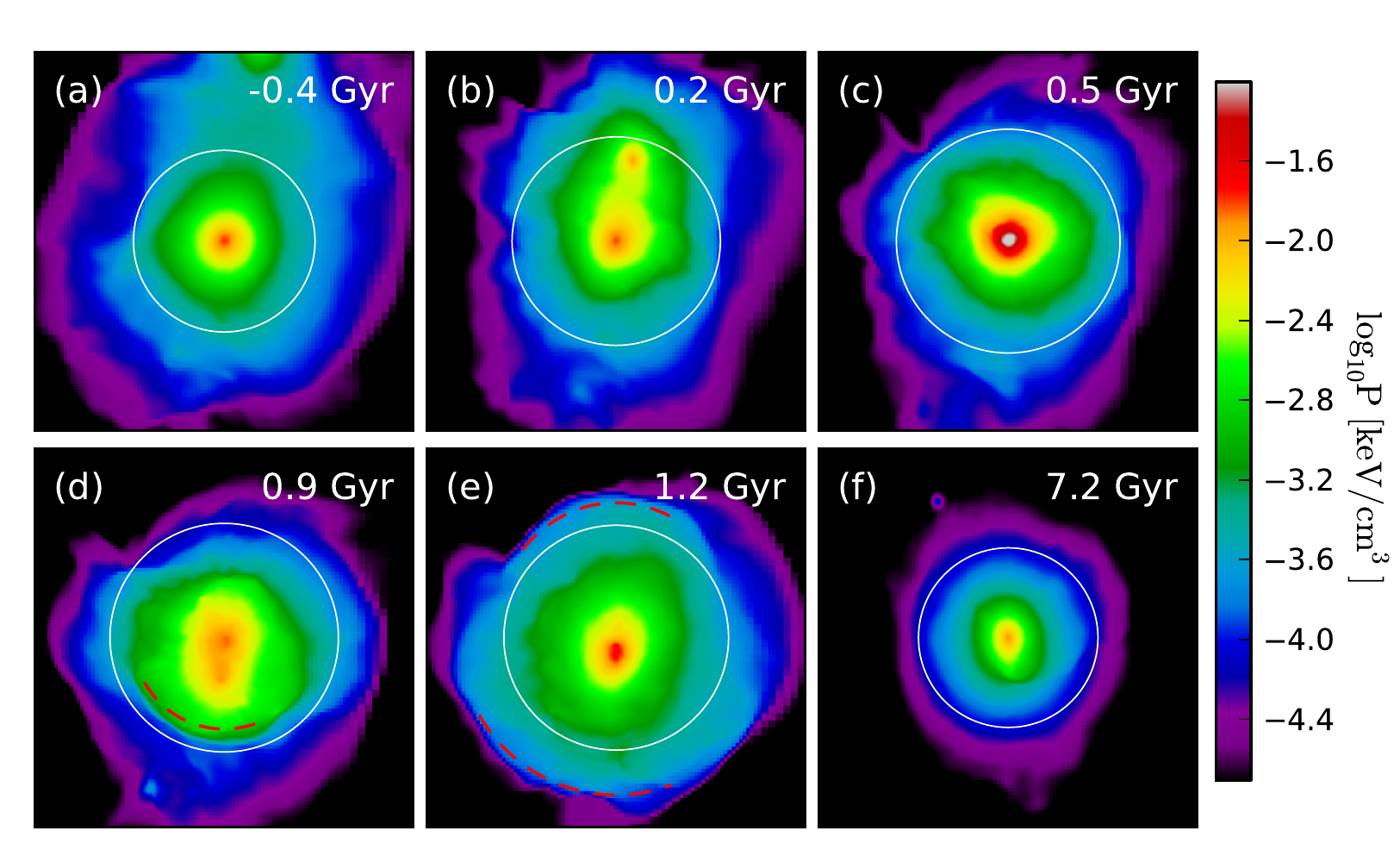}}

\caption{Thermal pressure maps of CL10 at six epochs surrounding the last major merger. Each panel is 2$h^{-1}$ Mpc (comoving) and 1$h^{-1}$ Mpc (comoving) along the line of sight, centered on the most massive progenitor. The white circles depict the size of $\rc$ and the red dashed arcs mark the location of the merger shock. The time relative to the beginning of the merger (the epoch when the merging cluster cross $\trc$) is shown in each frame. Frame (b) shows the epoch when the merging cluster crosses $\rc$, frame (c) shows the time of closest approach, and frame (f) shows a time when CL10 has relaxed significantly after the merger.}
\label{fig:images}
%\end{figure*}

%\begin{figure*}
\epsscale{0.35}

\centerline{\plotone{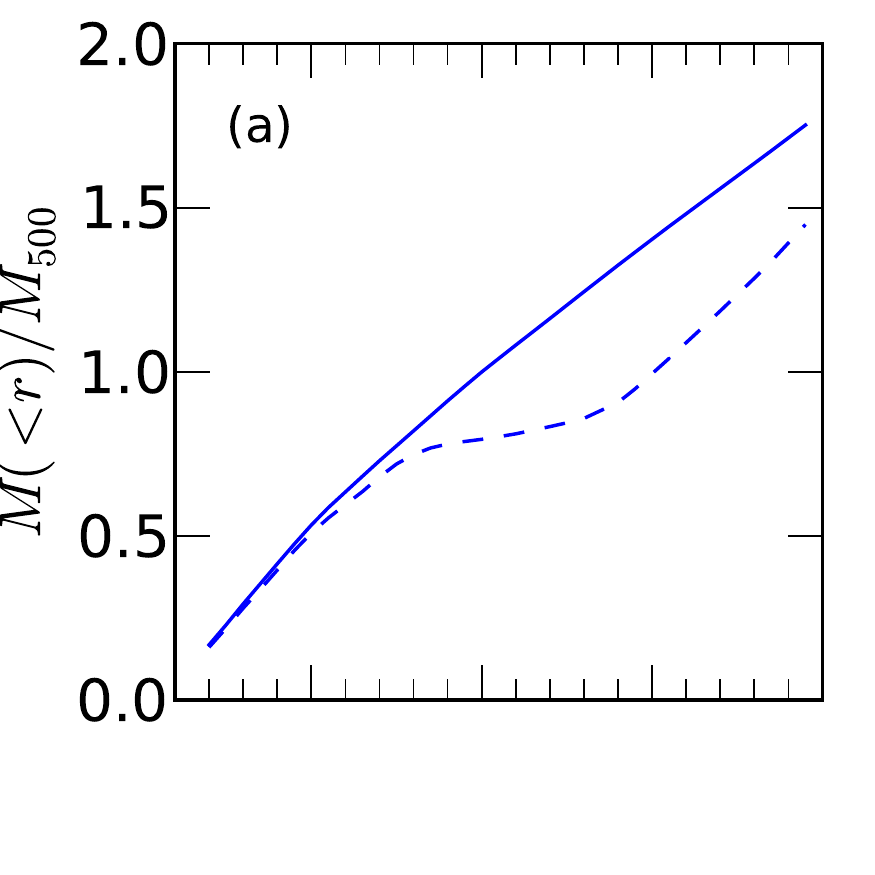}
		\hspace{-0.1in}\plotone{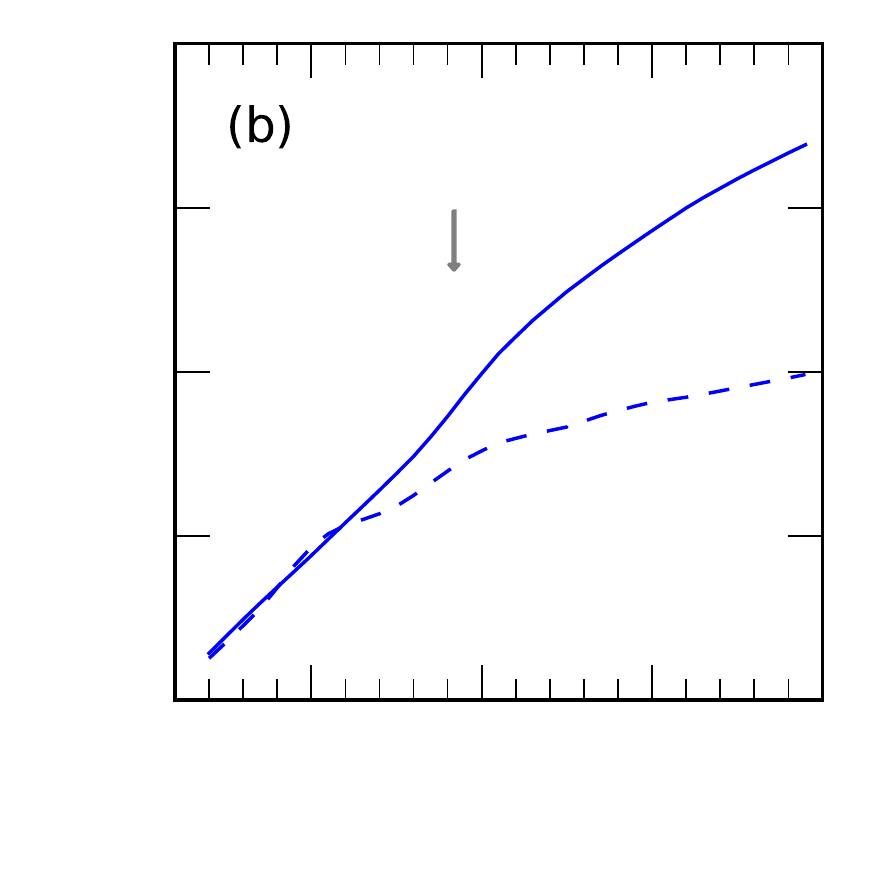}
		\hspace{-0.1in}\plotone{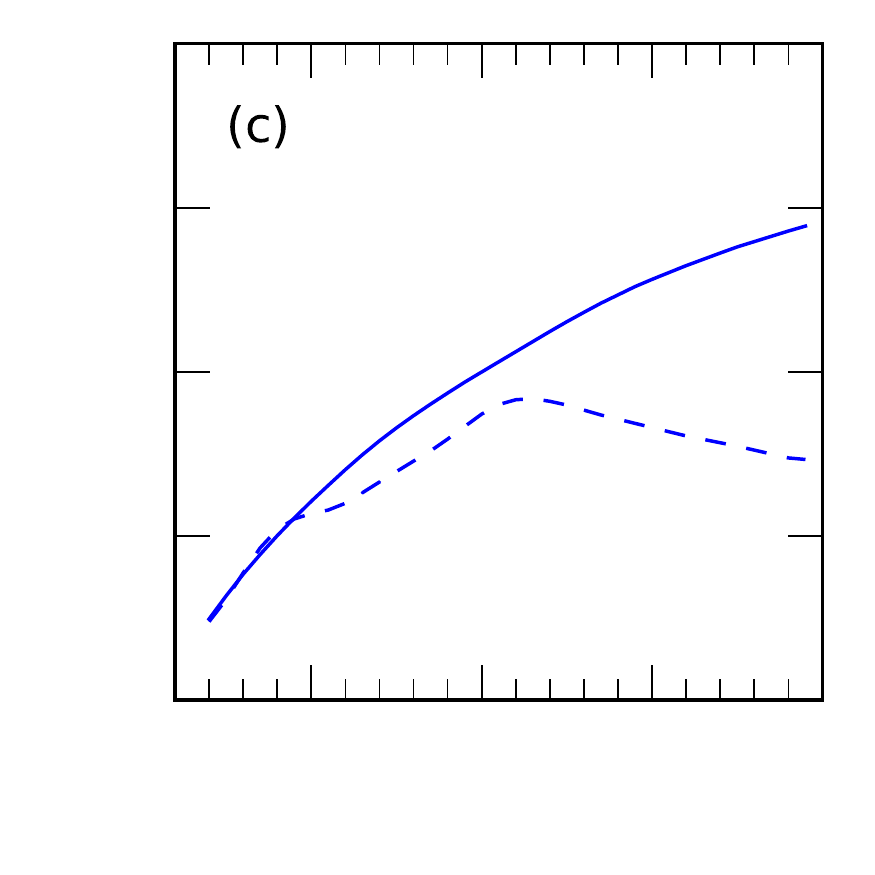}}
\vspace{-0.3in}
\centerline{\plotone{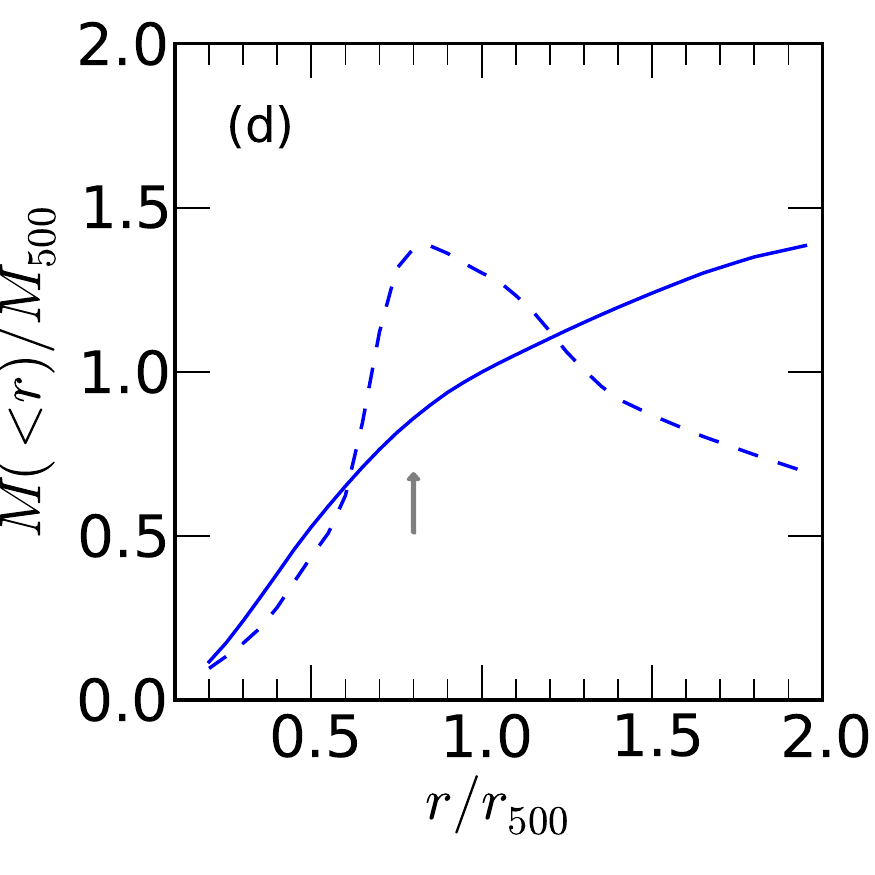}
		\hspace{-0.1in}\plotone{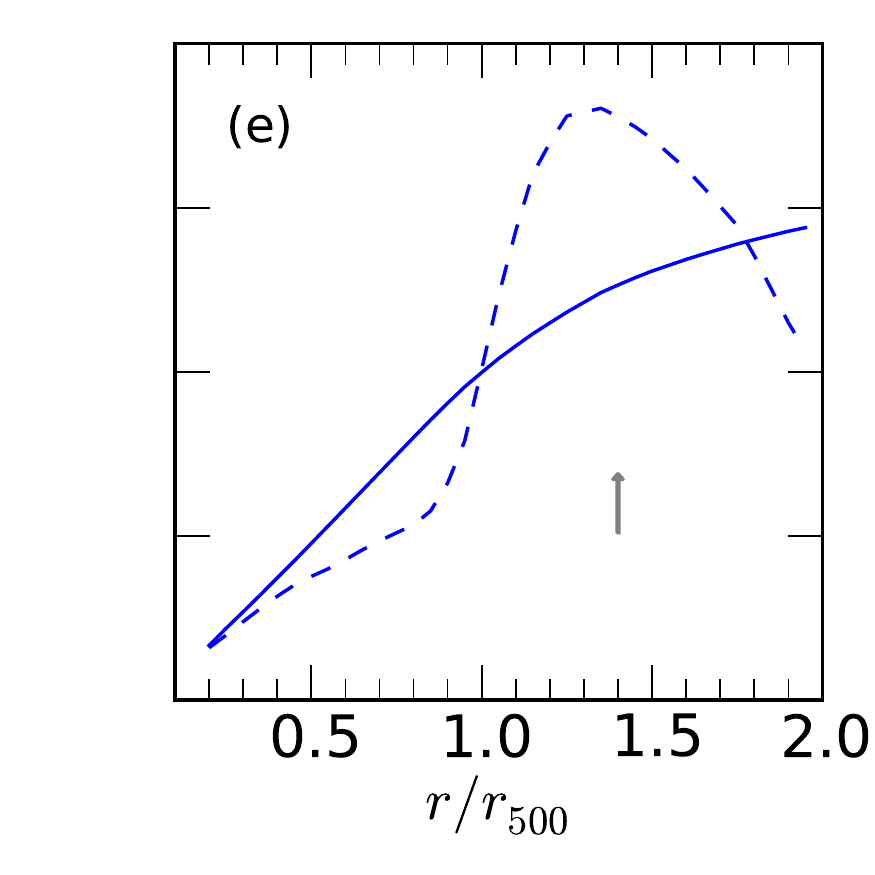}
		\hspace{-0.1in}\plotone{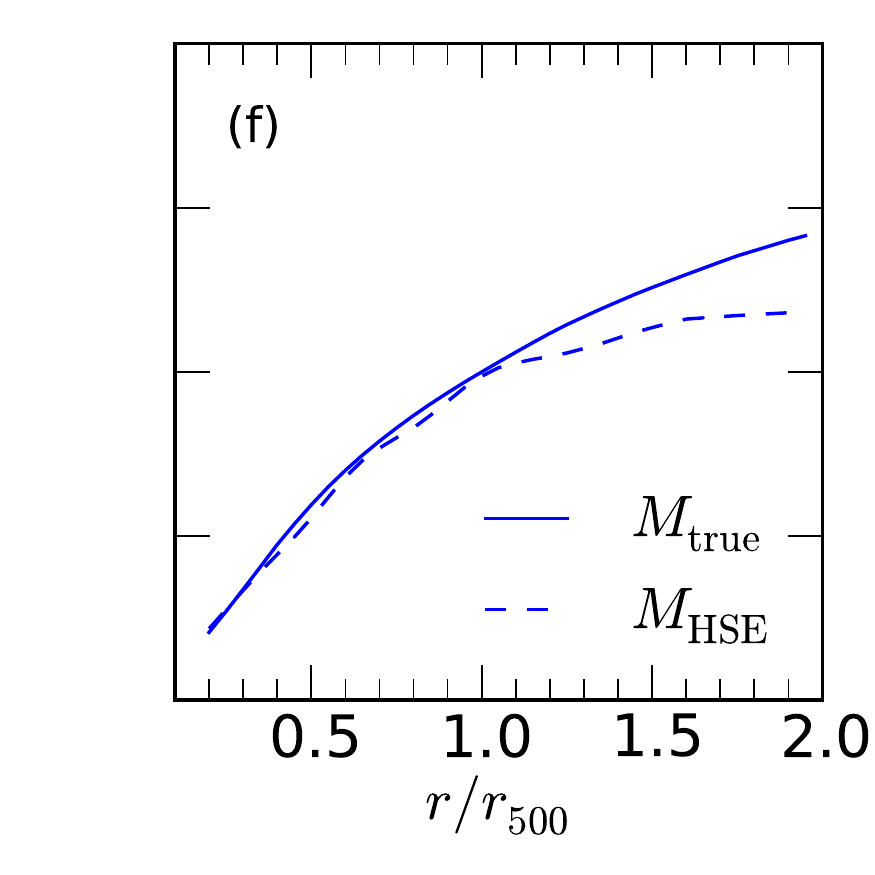}}

\caption{The mass profile of CL10 at the same six epochs depicted in Figure~\ref{fig:images}.  The hydrostatic mass estimate, $M_{\mathrm{HSE}}$, and the true mass, $M_{\mathrm{true}}$, are shown by the dashed and solid lines, respectively. The downward facing arrow marks the location of the incoming merging halo. The upward facing arrow marks the location of the shock wave created by the merger as it emanates outwards through the parent halo. }
\label{fig:multipanel}
\epsscale{1.0}
\end{figure*}

We begin  by exploring in detail the evolution of a single object, CL10, with a 
relatively quiescent merger history prior to a major merger at $z \approx 0.72$. This allows us to
isolate the effect of a single merger and compare directly to previous studies that explore
idealized binary mergers \citep[e.g.,][]{Ricker:2001,Poole:2006}.

%%%%% Discussion of the pressure maps

The response of the ICM thermal pressure during the merger is illustrated by Figure~\ref{fig:images}. Each frame is 2$h^{-1}$ Mpc comoving across and 1$h^{-1}$ Mpc comoving deep in projection. $\rc$ is marked with a thin white circle at each epoch, and $\trc \simeq 1.5\rc$ and $\tfrc \simeq 0.5\rc$. 
Initially CL10 is very relaxed, not having experienced a merger for 4.2 Gyr. Note the nearly circular
pressure contours within $\rc$ in the first panel.  The merging cluster, with mass ratio of 1:2, enters 
from the top of the frame with an orbit nearly in the plane of the projection, achieving closest approach 
approximately $\sim0.5$ Gyr after the start of the merger.  This naturally divides the merger into three epochs. 
In the initial passage (panels a--c) the spherically averaged outer pressure profile becomes shallower due to the presence of the 
merging cluster and compression of the ICM between the two cluster cores.  At the time of closest approach
the inner pressure profile becomes strongly peaked, generating forward and reverse merger shocks, marked with a dashed red arc
\citep{Ricker:2001}. In this second phase of the merger (panels d and e), the merger 
shocks propagate outward, crossing $\rc$ at $\sim0.9$ Gyr following the start of the merger.  After 
several sound crossing times the merger shocks have passed out of the cluster central region and merge 
with the large-scale accretion shock. The last panel of Figure~\ref{fig:images} (f) shows CL10 at $z = 0$, 
$\sim7.2$ Gyr after the merger. By this time the disruption of the merger has dissipated, leaving a 
relaxed, symmetric pressure profile similar to that seen at the start of the merger.

%%%%% Discussion of evolution of the profiles

Deviations from hydrostatic equilibrium can be easily understood given
the features seen in the pressure maps.  In
Figure~\ref{fig:multipanel}, we compare the true mass profile (solid
lines) to that predicted from the equation of hydrostatic equilibrium
(Equation~\ref{eq:hse}, dashed lines) at the same six epochs depicted
in Figure~\ref{fig:images}. The start and end states are again
qualitatively similar, with a cluster core in near-equilibrium and
progressively greater deviation with increasing radius beyond $\rc$.
The systematic underestimate of the total mass for relaxed clusters
has been noted previously \citep{Rasia:2004,Nagai:2007a,Lau:2009}.

The hydrostatic mass bias shows significant
evolution in both radius and time.  As the merging cluster makes its
initial passage (panel b), with position denoted by the downward arrow, the true
mass profile increases faster than the mass profile derived from the
thermal pressure, temporarily increasing the (negative) hydrostatic
mass bias at $\rc$.  

The outwardly moving merger shocks, with approximate positions shown by upward arrows in panels (d) and (e),
produce locally very steep pressure gradients, and thus strong overestimates of the true mass.
This is seen as a positive peak in the hydrostatic mass bias profile that propagates outward with the shocks.
The width of these peaks is caused by a combination of spherical averaging and the smoothing
procedure described in Section~\ref{sec:simulations}. This evolution is in agreement
 with the evolution of the hydrostatic mass profile 
during a merging event shown in \citet{Puchwein:2007}. 

The evolution of the hydrostatic mass bias at a fixed radius is
governed by the radial migration of these features across that radius.
Broadly, the bias will decrease slightly during the initial phase of
the merger. As the merger shocks move outward the bias will see a
dramatic rise, peaking at shock crossing before falling once more to a
large negative bias.  This bias will diminish at all radii as the
cluster continues to relax from the inside-out, however, at $r \gtrsim
\rc$ the bias persists to late times. As we will show in the next
section, these features seen in the mass bias for the CL10 merger are
generic to all the cluster mergers in our sample.

\begin{figure*}
\begin{center}
\epsscale{0.30}
\plotone{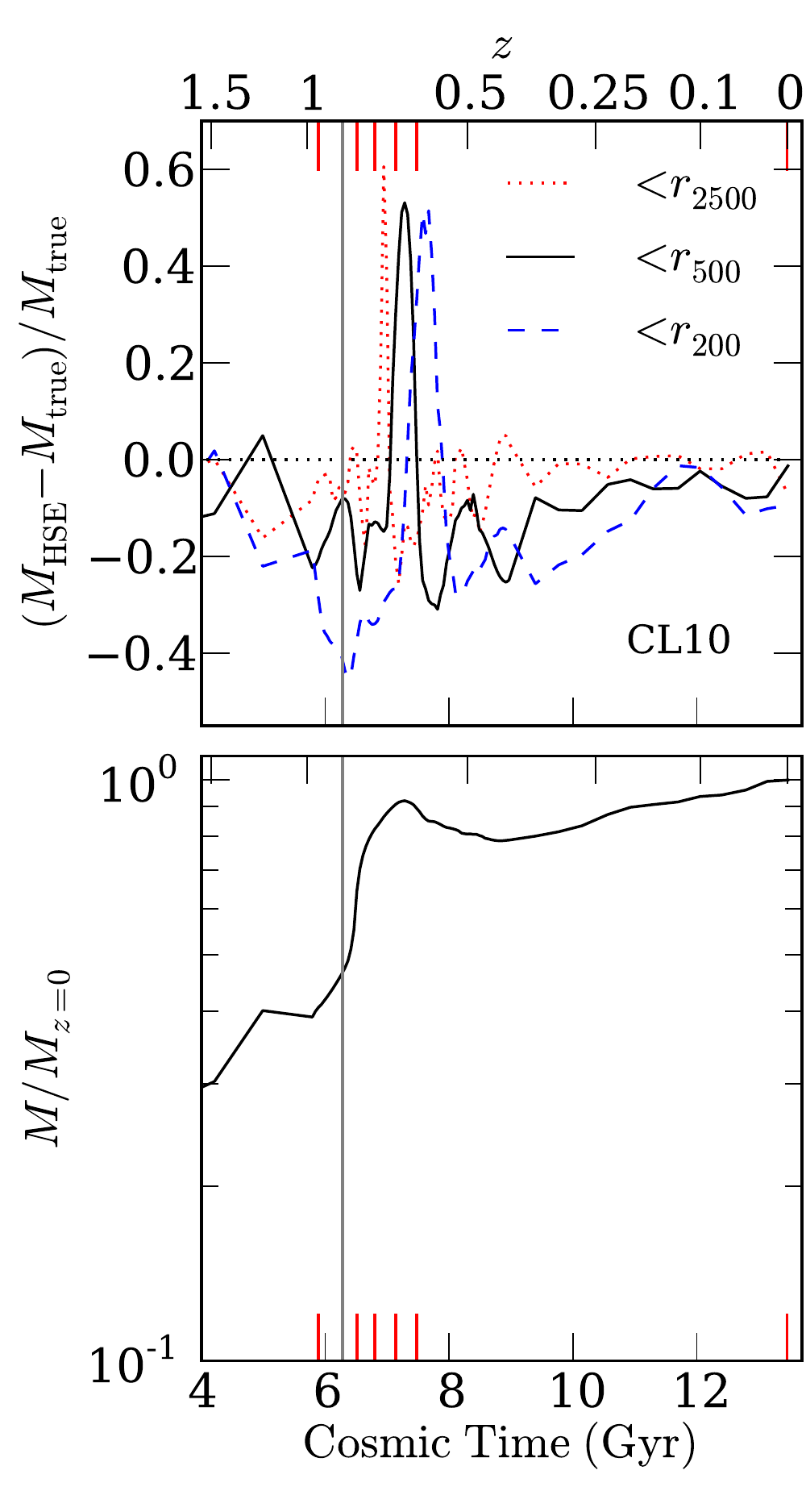}
\plotone{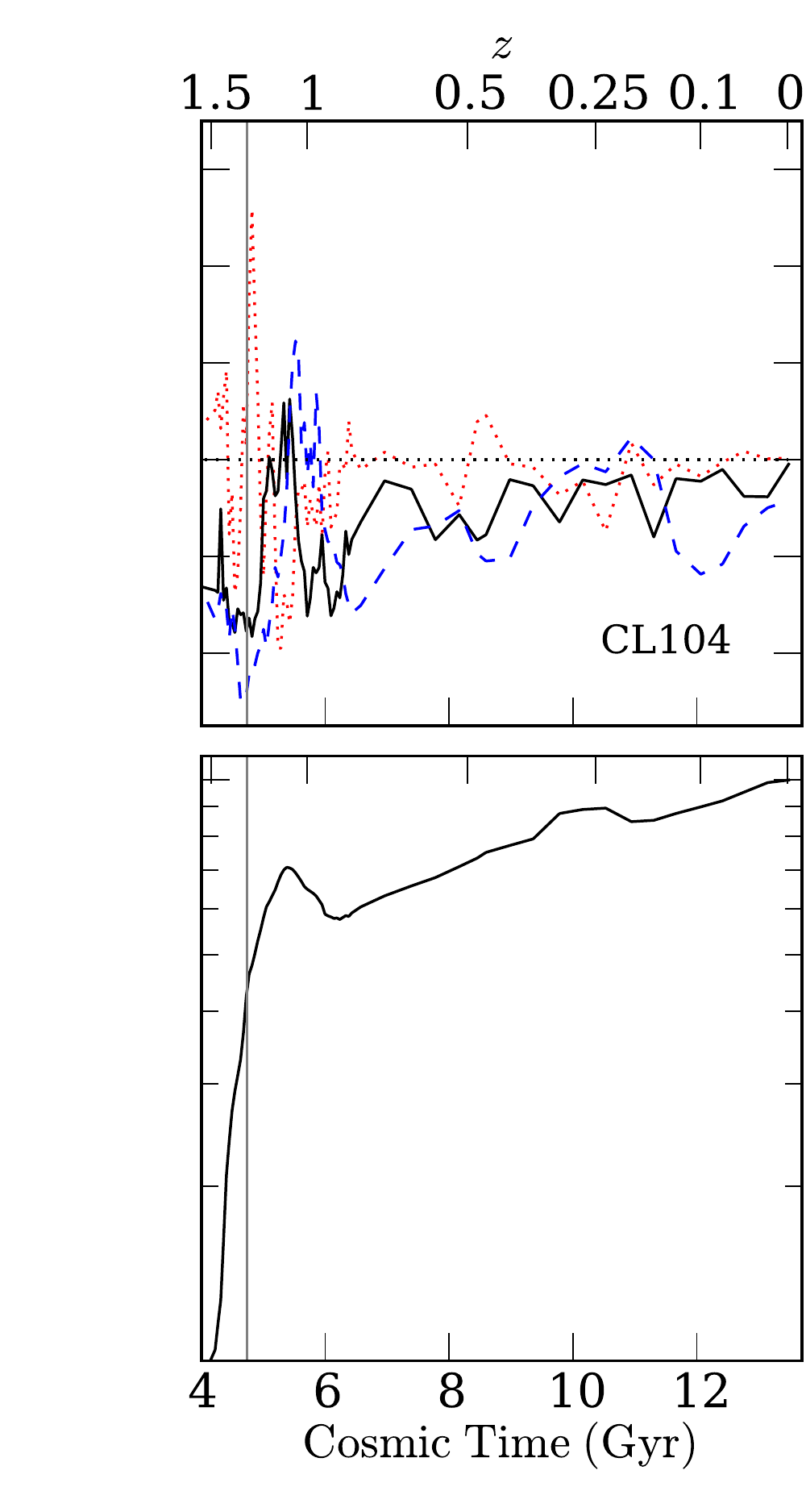}
\plotone{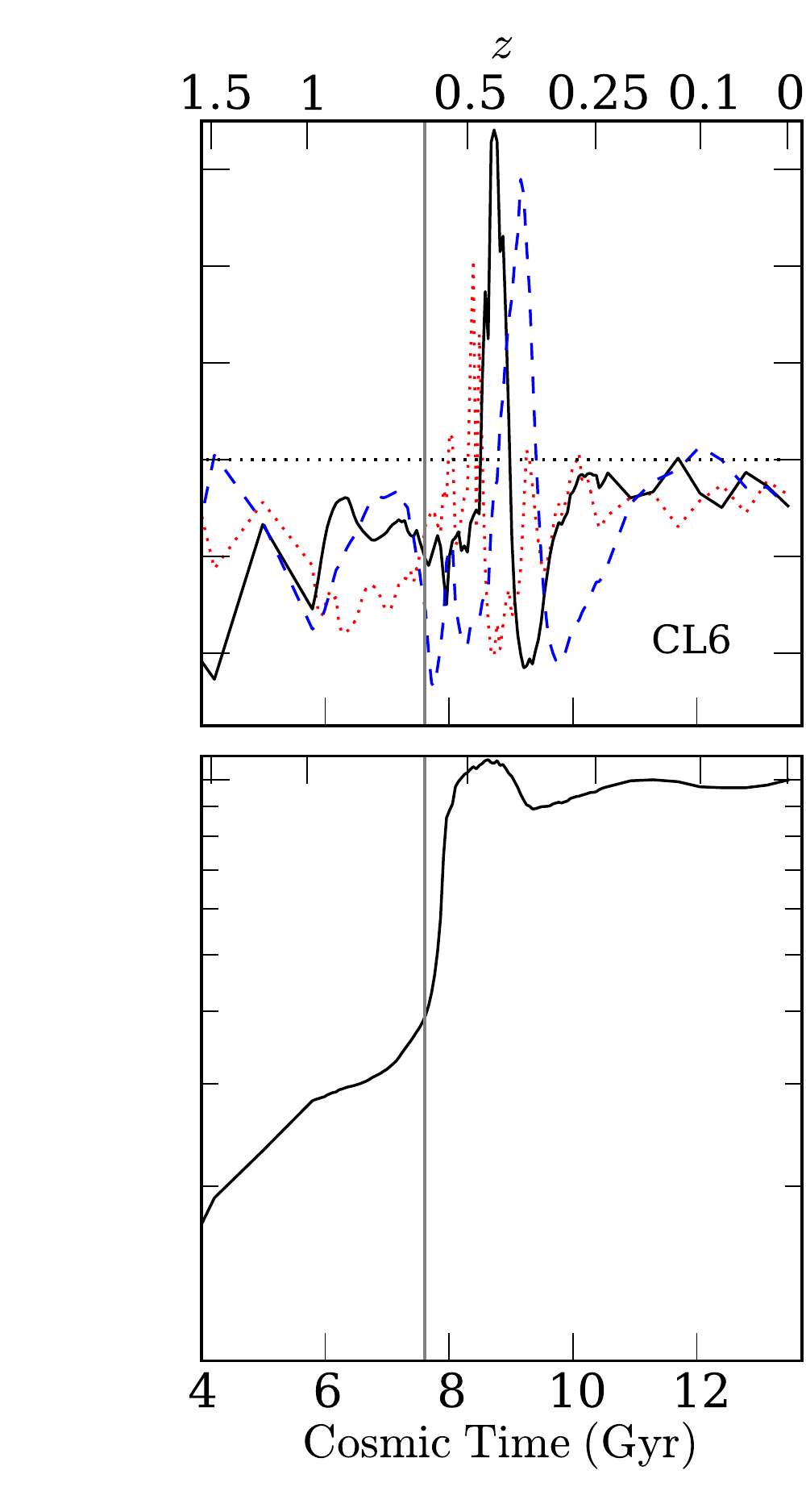}
\end{center}
\caption{Top: The time evolution of the mass bias (top) for three clusters, CL10, CL104, and CL6 at three radii $\tfrc$ (red, dotted), $\rc$ (black, solid),  and $\trc$ (blue, dashed). Bottom: The mass accretion history of the three clusters at $\rc$,  normalized by $M_{500}$ at $z = 0$. The vertical gray line marks the beginning of the latest major merger for each of the clusters. The red ticks mark the epochs in CL10 corresponding to the panels in Figures 1 and 2.}
\label{fig:bM}
\end{figure*}

We now expand our exploration of the hydrostatic mass bias beyond the
time immediately surrounding the CL10 merger.  The top left panel of
Figure~\ref{fig:bM} shows the evolution in the mass bias at 3
different radii, $\tfrc$, $\rc$, and $\trc$, chosen for their
relevance to X-ray and SZ observations. The bottom panel shows the
mass accretion history at $\rc$ normalized by the $M_{500}$ at $z = 0$. 
A vertical line marks the beginning of the final major merger. The epochs 
shown in the previous figures are marked with red tick marks along the 
x-axes of the panels.

The final major merger (depicted in the previous figures) begins at $\sim$6
Gyr. The mass of the cluster rises quickly as the incoming cluster
merges with the parent cluster. About 1 Gyr after $\trc$ crossing, we
see another sharp, very positive bias in the top panel. The bias falls
again and remains negative through $z$ = 0, oscillating slightly as
the cluster relaxes.

The evolution in the other radii is qualitatively similar, also
showing the short-lived, positive bias immediately following the
merger.  Note that the peak at each radius occurs at a different time
following the merger, with a delay of approximately $\sim0.5\Gyr$
between $\tfrc$ and $\rc$ and between $\rc$ and $\trc$, reflecting the
shock propagation we observed in panels (d) and (e) of Figure~\ref{fig:multipanel}.
Immediately following the peak, the hydrostatic mass again
underestimates the true mass and this negative bias holds but steadily
diminishes to $z = 0$.

For comparison we also show in Figure~\ref{fig:bM} the evolution of
two clusters selected from our full sample of 16. CL104 is a massive
cluster that forms from a complicated triple merger at $z \approx 1.5$
and remains mostly quiescent thereafter.  CL6 is intermediate in mass between
CL10 and CL104, and experiences a nearly one-to-one merger at $z \approx
0.6$.  Despite the differences in merger history and $z = 0$ mass, the
clusters show remarkable similarity in their evolution. All three
clusters have peaks $\sim$1 Gyr following the identified merger, and
show similar relative offsets between the different radii.  

Cluster CL104 shows less prominent merger characteristics than seen in
the other two clusters due to the triple merger, resulting in broader 
and smaller amplitude peaks at each radius. Following the major merger
at $z \approx 1.5$, CL104 also undergoes a minor merger at $z \approx 0.25$. However, this merger
only has a mass ratio of 10:1. Consequently the minor
merger does not impact the cluster enough to have a significant effect of the mass estimate.

By $z = 0$, the mass bias for all three clusters approaches zero. However, it is important 
to note that even at $z = 0$ the hydrostatic mass bias within $\rc$ is still biased low by
5--10\% in the outskirts.

\subsection{General Trends in the Hydrostatic Mass Bias}
\label{sec:averagedhse}

\begin{figure}[t]
\begin{center}
%\epsscale{0.40}
\plotone{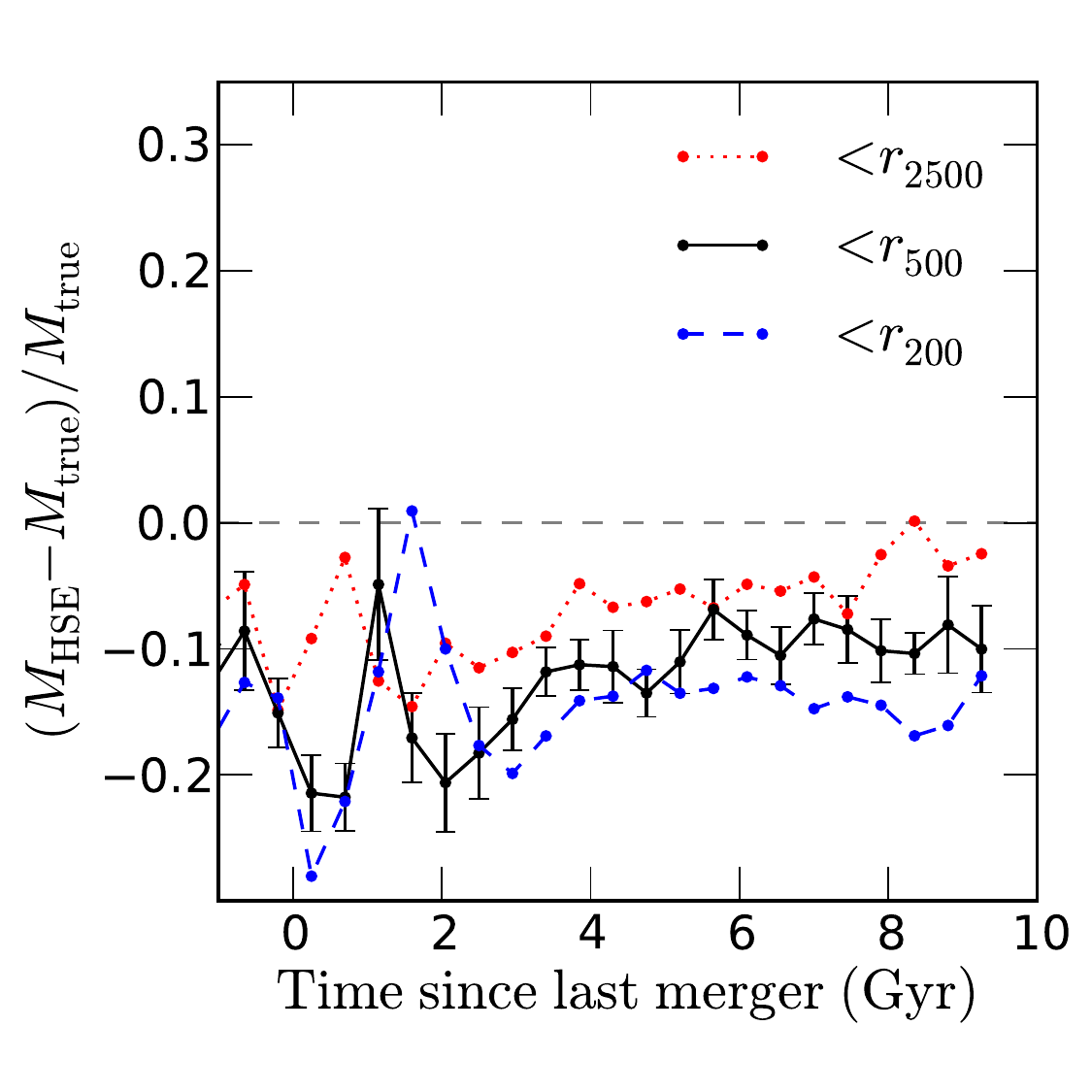}
\caption{Averaged mass bias as a function of time elapsed since last merger (in Gyrs) for the sixteen clusters. A more detailed discussion of this figure is located in \S \ref{sec:averagedhse}. The biases are plotted at radii $\tfrc$ (red, dotted), $\rc$ (black, solid),  and $\trc$ (blue, dashed).  The error bars show the $1\sigma$ error on the mean at $\rc$.}
\label{fig:bMtmerger}
\end{center}
\end{figure}

The post-merger evolution in Figure~\ref{fig:bM}
exists for all 16 clusters in our simulated sample.  This is shown
explicitly in Figure~\ref{fig:bMtmerger}, where the
average evolution in the hydrostatic mass bias is plotted for the
entire sample.  The final major merger is identified for each
cluster (see Table~\ref{table:clusters}) and used to define a common
point between clusters with varied accretion histories. The hydrostatic 
mass bias is measured at three separate radii, $\tfrc$, $\rc$, and $\trc$, 
linearly interpolated to equally spaced times relative to the last
major merger, and averaged over the entire sample.  Error bars
denote the error on the mean for $\rc$. The error bars for $\tfrc$ 
and $\trc$ are comparable, and omitted for clarity.  Note that clusters 
that undergo low-redshift mergers will only contribute to the 
low-$t_{\mathrm{merger}}$ portion of the figure. 
Consequently, while the behavior exhibited in the 
early epochs of Figure~\ref{fig:bMtmerger} is an average over the entire
sample, the later epochs are necessarily biased to a small number of
early forming objects.  We plot the evolution of the mass bias to 9.25 Gyr 
following the last merger at which point fewer than three clusters contribute 
to the average.

The qualitative features for the individual clusters presented in
Figure~\ref{fig:bM} are also apparent in the sample-averaged
evolution in Figure~\ref{fig:bMtmerger}.  Immediately preceding the 
merger, the hydrostatic mass underestimates the true mass by $\approx10\%$ 
at all radii. This bias grows to nearly $30\%$ in the outskirts 
shortly after the merger begins. Following the peaks associated with 
the merger shocks, the hydrostatic mass bias decreases over the 
subsequent 8 Gyr from $\sim$15--20\% to 3\%, 10\% and 18\% within 
$\tfrc$, $\rc$ and $\trc$, respectively.

The peaks caused by the propagating merger shocks have smaller
amplitude and are broader than the individual cluster histories seen
in Figure~\ref{fig:bM}. Therefore, the mean mass bias remains negative at
all $t_{\rm merger}$. This is caused by a variety of factors,
including differences in the merger mass ratio and the orbital impact
parameter, which changes the time offset between the peak and the
start of the merger.  Since the peaks are narrow in time (typically
less than 0.5 Gyr), and we are primarily concerned with the average
evolution after the merger, we make no attempt to reduce this
broadening.

We also examined how the mass bias depended on various parameters,
such as merger ratio, final cluster mass and redshift. We found little
dependence on present mass of the cluster, with the larger clusters
experiencing a slightly greater peak in the mass bias immediately
following merger.  

\section{Influence of Non-Thermal Pressure on the Hydrostatic Mass Bias}
\label{sec:pturb}

\begin{figure*}%[t]
\epsscale{0.90}
\plottwo{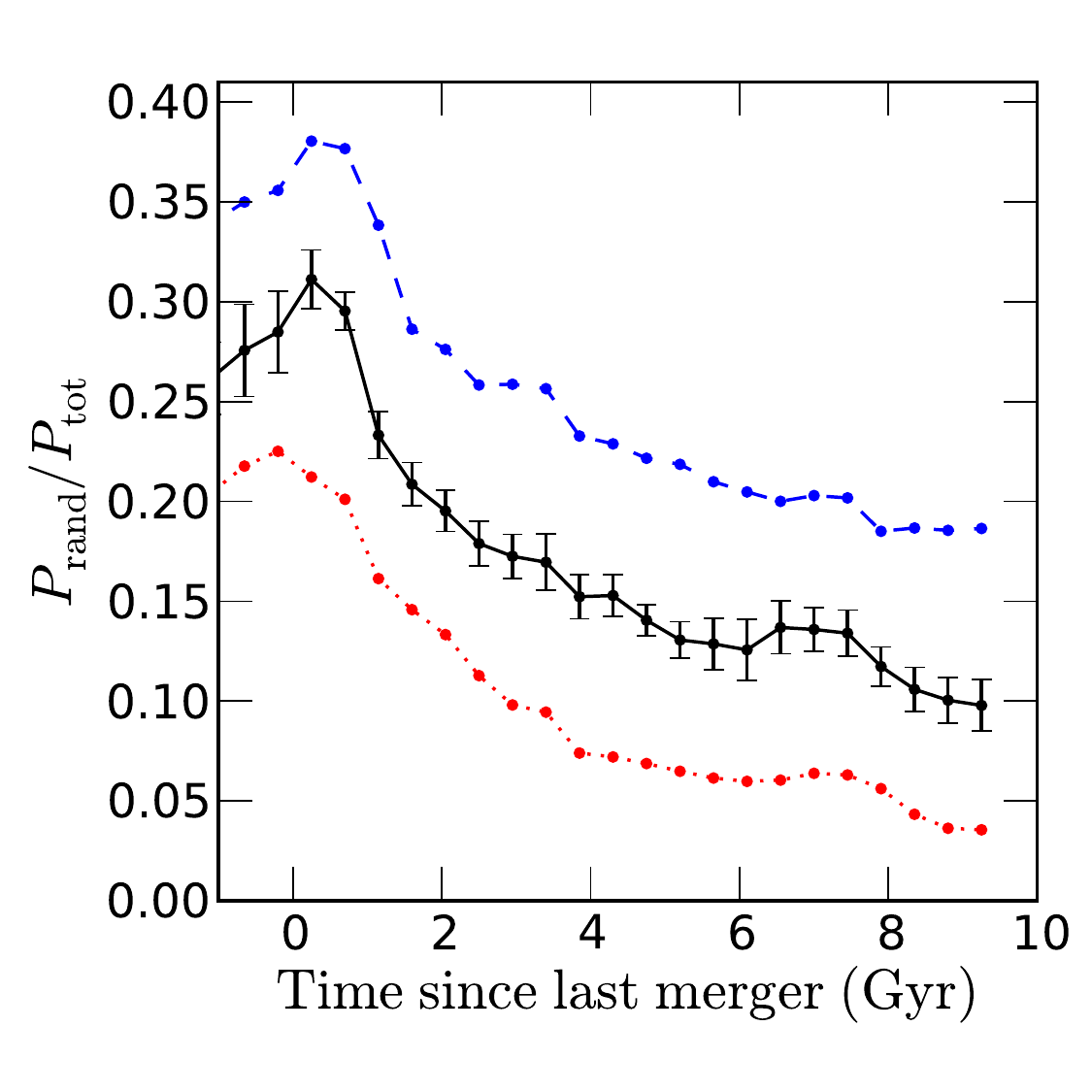}{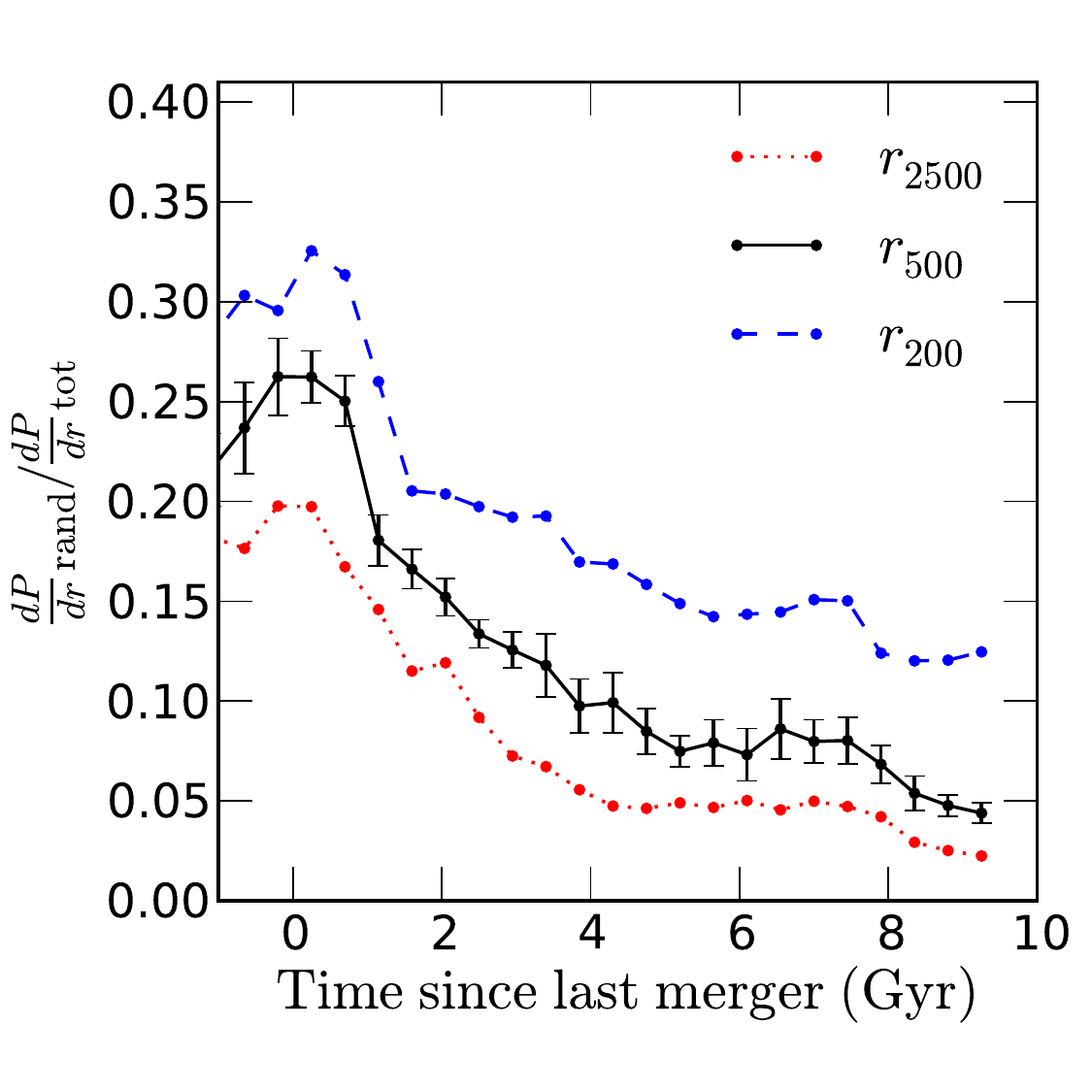}
%\vspace{-0.2in}
\plottwo{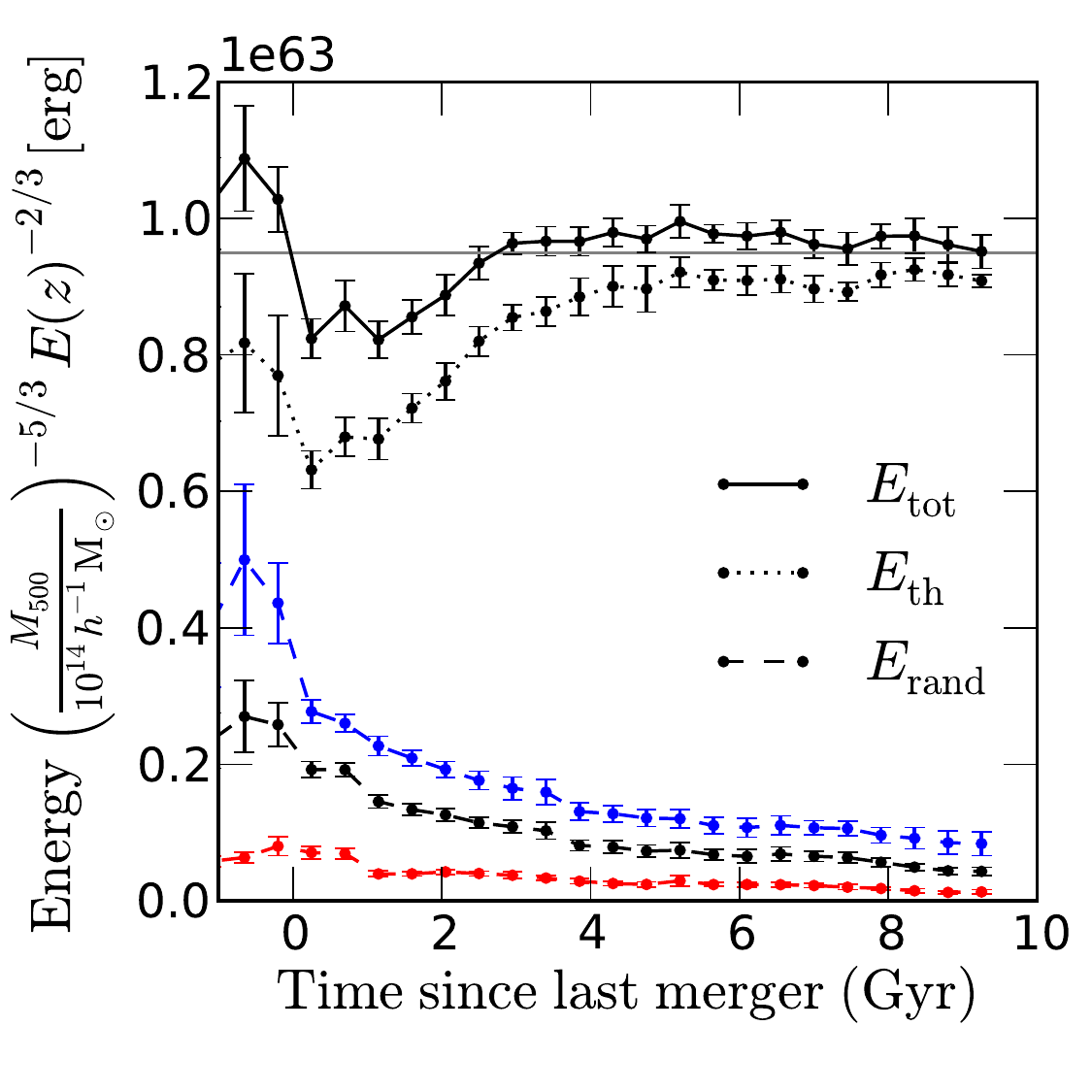}{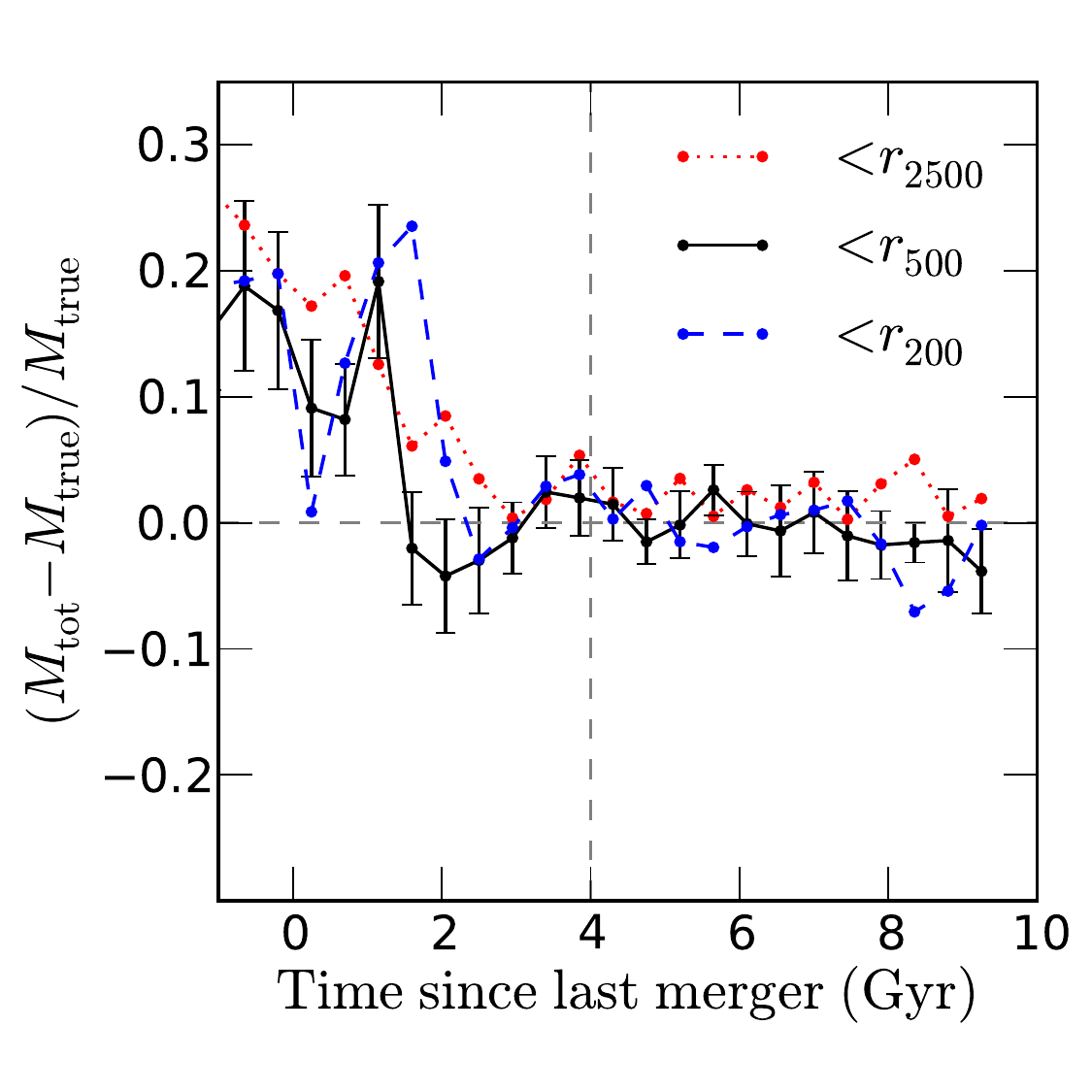}
\caption{\textit{Top}: Fractional contribution of random motions to the total effective pressure 
(left) and pressure gradient (right). \textit{Bottom left}: The components of the total ICM energy. \textit{Bottom right}:    Averaged corrected mass bias as a function of time elapsed since last merger (in Gyrs) for the sixteen clusters.  In all figures, the quantities are plotted at radii $\tfrc$ (red, dotted), $\rc$ (black, solid),  and $\trc$ (blue, dashed). Our sample is divided into relaxed and unrelaxed clusters at $t_{\rm merger}$ = 4 Gyr (marked by the vertical dashed line). The error bars show the $1\sigma$ error on the mean at $\rc$.}
\label{fig:pturb_diff}
\end{figure*}

%%%%% Discussion of differential turbulence

Figures~\ref{fig:bM} and \ref{fig:bMtmerger} demonstrate that the
magnitude of the mass bias decreases steadily following a merger,
yet never fully disappears.  If, as has been previously suggested
\citep{Lau:2009,Nagai:2007a,Rasia:2004}, residual motions in the ICM 
account for this bias, the non-thermal component of pressure due
to these motions should see a similar evolution to the hydrostatic
mass bias. We now investigate the evolution of this non-thermal 
component in more detail.

The top panels of Figure~\ref{fig:pturb_diff} show the evolving
contribution of random motions to the total ICM pressure (left panel)
and pressure gradient (right panel) averaged over all 16 clusters in 
the same manner as in Figure~\ref{fig:bMtmerger}. The measurement of 
the non-thermal pressure support is described in Section~\ref{sec:clusterprofiles}.  
During the merger, the effective pressure from random motions
is $\sim30\%$ at $\rc$ and has a strong trend with radius, increasing
from $\sim22\%$ at $\tfrc$ to $\sim38\%$ at $\trc$. This radial trend 
persists throughout the merger and during the following
relaxation. The non-thermal pressure fraction drops rapidly over the next 
1--2$\Gyr$, and steadily decreases thereafter from $\sim20\%$ to $\sim10\%$ at
$\rc$.  

The radial pressure gradient shown in the top right panel of Fig.~\ref{fig:pturb_diff}, an analog to the thermal pressure gradient in the equation of hydrostatic equilibrium (Eq.~\ref{eq:hse}), shows similar trends in both radius and time since merger. 
Initially, random velocities contribute $\sim20\%$, $\sim26\%$, and $\sim32\%$
to the non-thermal pressure gradient for $\tfrc$, $\rc$, and $\trc$, 
respectively.  This falls to between $2-12\%$ after $8 \Gyr$, in line
with the residual hydrostatic mass bias at those radii.

Our findings are consistent with the measurements of previous work. \citet{Lau:2009} 
also sees small scatter in the gradient of non-thermal pressure at $\rc$ from 4--10\%
among clusters identified as relaxed at $z = 0$ and a larger range of
3--18\% for unrelaxed clusters (see their Figure 3). While \citet{Lau:2009} used the same simulations as this work, our mass biases are also in general agreement with the measurements from other simulations, e.g. GADGET-2 in \citet{Rasia:2006,Piffaretti:2008,Battaglia:2011a}.  The strong radial dependence in the non-thermal pressure fraction, with larger non-thermal support at greater radii is consistent with the results by \citet{Iapichino:2008},  \citet{Battaglia:2011a}, and  \citet{Vazza:2011}. In addition, the dependence on dynamical state seen in top panels of Figure~\ref{fig:pturb_diff} is consistent with the findings of \citet{Vazza:2011}. They also demonstrate that clusters undergoing mergers have a significantly larger fraction of non-thermal pressure support than relaxed or relaxing systems.

%%%%% Discussion of cumulative turbulence

We expect the decay of the non-thermal component seen in the upper panels of
Figure~\ref{fig:pturb_diff} is due to the
conversion of random velocities to thermal energy via shocks and
numerical dissipation (no explicit physical viscosity was included in
our simulations).  Demonstrating this explicitly is difficult, however,
as the total energy in the ICM increases by continued accretion subsequent
to the merger, as can be seen in the bottom panels of Figure~\ref{fig:bM}.
To account for this growth we rescale the energy by the self-similar 
mass and redshift scaling $M^{5/3} E(z)^{2/3}$, where $E(z) = H(z)/H_0$ is the 
dimensionless Hubble constant \citep{Kaiser:1986,Nagai:2007b}.
In the bottom left panel of Figure~\ref{fig:pturb_diff} we show the
the evolution of the total (solid), thermal (dotted) and
random kinetic (dashed) energy of the ICM within $\rc$, scaled by the
self-similar expectation. To show the evolution of energy at different 
radii, we also plot the random kinetic energy within three radii: 
$\tfrc$ (red), $\rc$ (black), and $\trc$ (blue). 

The total energy content of the ICM remains essentially constant 
beyond the $2\Gyr$ period following the merger.  The time surrounding
the merger shows a significant decrease in $E_{\rm tot}$, however 
this is due not to a decrease in energy but rather the temporary
increase in $M_{500}$ caused by the presence of the merging cluster
in the cluster core.  Much of this mass will move to larger radius,
decreasing the mean density and thus $M_{500}$, as can be seen
in the accretion histories shown in Figure~\ref{fig:bM}.
The apparent decrease in energy is a reflection of this deviation
from self-similarity, rather than a change in the total ICM energy. 
Since the total energy remains constant, we can associate directly
the decrease in random kinetic energy with the increase in thermal 
energy.  This thermalization is in agreement with the
findings of \citet{Vazza:2011} who show that the fraction of turbulent
energy to thermal energy increases with decreasing time since major
merger. This evolution in the thermal energy has important implications 
for the behavior of cluster mass proxies such as the integrated SZ flux 
\citep[e.g.][]{Nagai:2006, Shaw:2008, Yang:2010} or X-ray temperature 
\citep[e.g][]{Ritchie:2002,Rowley:2004, Rasia:2011} which are sensitive 
to the thermal energy content of the ICM, and will be explored in greater 
detail in future work.

%%%% Discussion of corrected Mass

It is interesting to ask at what point the large bulk velocities associated with
the merger progenitors reach equilibrium with the new gravitational potential. 
We produce a corrected mass estimate, $M_{\rm{tot}}$, using 
Equation~\ref{eq:MJeans} and measuring the terms due to random motions and
rotation.  

The lower right panel of Figure~\ref{fig:pturb_diff} shows the evolution of $M_{\mathrm{tot}}$ 
following the merger.  The period up to $\approx2 \Gyr$ following the merger shows rapid 
evolution, with the mass bias ranging from a positive $20\%$ bias to a negative $10\%$ bias.
This is driven primarily by the evolution in $M_{\mathrm{HSE}}$ caused by the outwardly 
propagating merger shock seen in Figure~\ref{fig:images}.  
The mean M$_{\mathrm{tot}}$ after 4 Gyr is essentially unbiased, with a mean of $0.3\%$ and 
$\approx8\%$ scatter at $\rc$.  This indicates that the velocities at this radius have reached 
equilibrium with the cluster potential.

\subsection{Correcting the Hydrostatic Mass Bias at z = 0}

\begin{figure}[t]
\epsscale{1.15}
%\epsscale{0.5}
\plotone{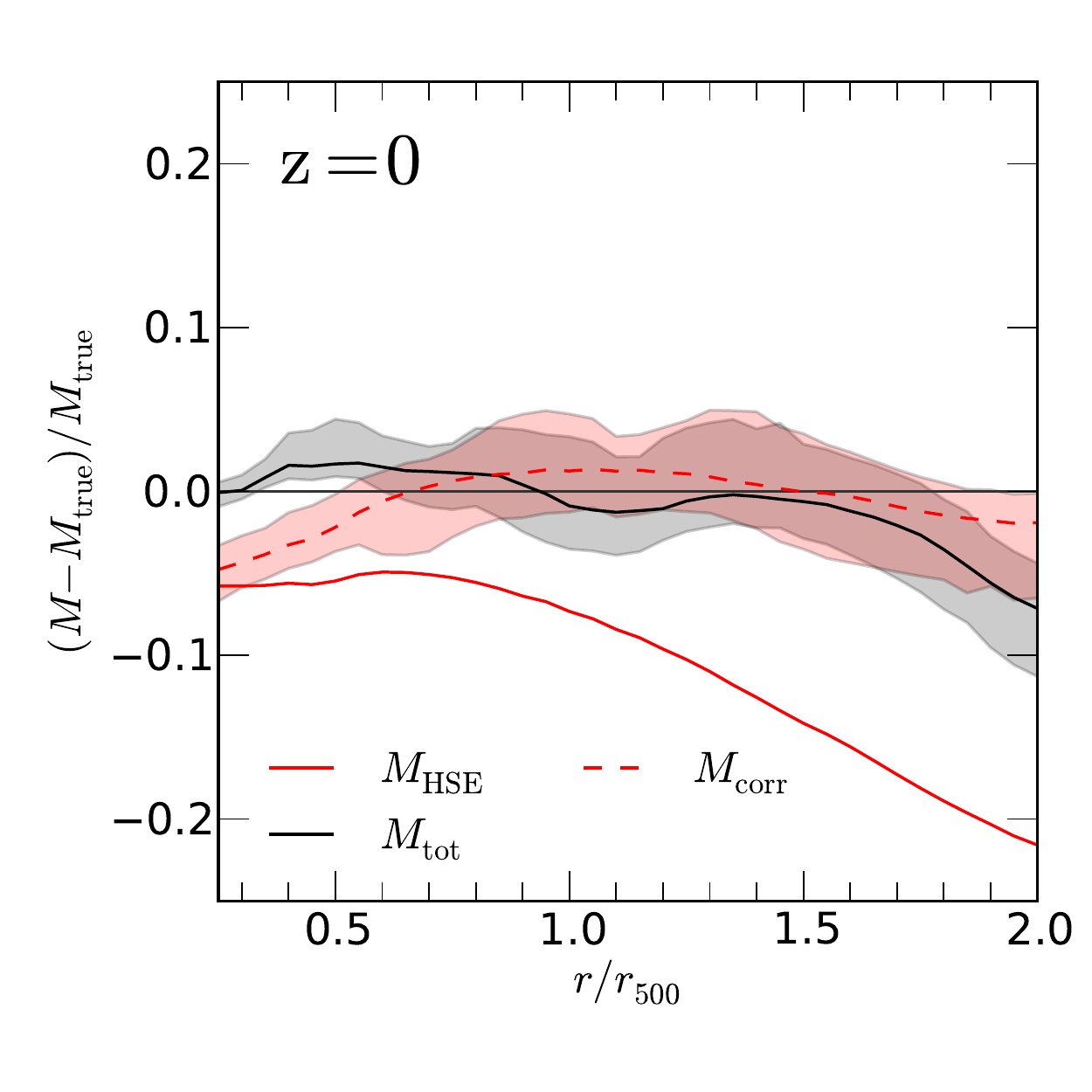}
\caption{The averaged mass measurements for the relaxed subsample, split at $t_{\mathrm{merger}}$ of 4 Gyr at $z$ = 0.  We show  $M_{\mathrm{tot}}$ (solid black)  and the mass corrected using the relation from \citet{Shaw:2010} fit to our subsample, $M_{\rm corr}$ (dashed red). See text for description of these mass calculations. $M_{\mathrm{HSE}}$ is shown with the solid red line for comparison.}\label{fig:McorrModel}
\end{figure}

The dependence of $M_{\mathrm{tot}}$ on the time since major merger
suggests a promising division between ``relaxed" and ``unrelaxed"
clusters may be obtained by splitting a sample between objects that
have and have not experienced a major merger within the past $4 \Gyr$. 
This results in a sample of 10 relaxed clusters at $z = 0$ 
(see Table~\ref{table:clusters}). Note that this includes all 6 clusters 
in Nagai et al. 07 identified as relaxed clusters based on morphological
classification.  Our relaxed subsample also contains an additional 4 
clusters that were missed by the morphological classification scheme. 
This suggests that these clusters, while dynamically relaxed, appear 
morphologically disturbed in the X-ray images.

Figure~\ref{fig:McorrModel} shows the average $M_{\mathrm{HSE}}$ (solid red line) and
$M_{\mathrm{tot}}$ (solid black line) radial profiles for our relaxed clusters at $z = 0$. 
The inclusion of $M_{\rm rot}$ and $M_{\rm rand}$ in the calculation of $M_{\mathrm{tot}}$ removes 
the bias that is apparent for $M_{\mathrm{HSE}}$ out to $\approx1.5 \rc$, with a slight increase in scatter from 8\%
to 11\% at $\rc$ and 9\% to 10\% at $\trc$. 

Since $P_{\rm{rand}}$ is not accessible observationally, we propose a model
based on the average $P_{\rm{rand}}$ profile measured in our simulations.   We neglect the contribution
from rotation, which is $\lesssim 2\%$ of the mass correction beyond $0.7 \rc$, and  also assume velocity anisotropy, $\beta =  1- \sigma_r^2/2\sigma_r^2$, is zero.  While $\beta$ is nonzero and increases radially, we find that the inclusion of $\beta$ as measured from the simulations has little effect on the accuracy of our model. We 
parameterize the fractional contribution of the non-thermal pressure as in \citet{Shaw:2010},

\begin{equation}\label{eq:prandfrac}
\frac{P_{\mathrm{rand}}}{P_{\rm{rand}}+P_{\rm{th}}} = \alpha(r/r_{500})^{n}.
\end{equation}
Using the average $P_{\rm rand}$ and $P_{\rm th}$ profiles for our relaxed sample and fitting
over the radial range $0.2 \le r/\rc \le 2.0$, we obtain best fit values $\alpha = 0.14 \pm 0.09$ and $n = 1.16 \pm 0.17$. Results are mildly dependent on redshift in the redshift range $z\lesssim 0.2$ \citep[see][for redshift evolution]{Shaw:2010,Battaglia:2011a}. These differ from the best fit values given in \citet{Shaw:2010} of $\alpha = 0.18$ and $n = 0.8$, which were derived for the complete sample, including unrelaxed clusters. The resulting corrected mass profile contains only observable quantities $\rho_{\rm{gas}}$, $P_{\rm{th}}$, and $dP_{\rm{th}}/dr$, 
and is given by
\begin{eqnarray}
M_{\mathrm{corr}} & = & \frac{-r^2}{G\rho_{\mathrm{gas}}} \biggl(\frac{1}{1-\alpha (r/r_{500})^n} \frac{dP_{\rm{th}}}{dr} \biggr. \hspace{0.65in} \nonumber \\ 
 & & \hspace{0.65in} + \biggl. \frac{\alpha n (r/r_{500})^{n-1}}{r_{500} (1-\alpha (r/r_{500})^n)^2} P_{\rm{th}} \biggr).
\end{eqnarray}

The dashed red line in Figure~\ref{fig:McorrModel} shows the average $M_{\rm{corr}}$ for the sample
of relaxed clusters.  Using the average non-thermal pressure profile we are able to recover the true mass
between 0.7--2$\rc$. At smaller radii, $r \lesssim 0.7\rc$, the neglected rotational component becomes 
important and the mass is again underestimated.  A comparison of the mean mass biases 
for $M_{\rm corr}$ and $M_{\rm HSE}$ at two radii are shown in Table~\ref{table:bias}.  
It is worth noting that the mean mass biases for both the full and relaxed subsample are in good 
agreement with those of \citet{Lau:2009}, despite differences in gas physics employed 
in the respective runs; i.e., non-radiative simulations in this work and the simulations including 
cool and star formation in \citet{Lau:2009}. The $M_{\rm corr}$ mass estimate is able to 
recover the true mass on average to within 1$\%$ with 10\% scatter. This is a significant improvement 
over the mean HSE mass bias as measured for the full sample of 5--10\% with 15--20\% scatter. 

%*********************************************************************
\begin{deluxetable}{llrr}
\tablecolumns{3}
%\tablewidth{\linewidth} 
\tablecaption{ Mean Biases in Mass Estimates at z=0 \label{table:bias}}
\tablehead{
 & & \multicolumn{2}{c}{Bias~$\pm$~Scatter\tablenotemark{2}} \\
\cline{3-4}\\[-1.7ex]
{Mass Estimate} & Sample\tablenotemark{1} &  \multicolumn{1}{c}{ $\rc$} & \multicolumn{1}{c}{$\trc$}
}
\startdata
$M_{\rm HSE}$ & Full & $-0.10 \pm 0.16$ & $-0.06 \pm 0.19$  \\
$M_{\rm HSE}$ & Relaxed & $-0.07 \pm 0.08$ & $-0.14 \pm 0.09$ \\
$M_{\rm tot}$ & Relaxed &  $-0.01 \pm 0.11$ & $-0.01 \pm 0.10$ \\
$M_{\rm corr}$ & Relaxed & $0.01 \pm 0.09$ & $0.00 \pm 0.11$ 
\enddata
\tablenotetext{1}{Relaxed clusters are clusters that have not experienced a major merger in at least 4 Gyr.}
\tablenotetext{2}{The error on the mean is given by the scatter divided by $\sqrt{N-1}$, where $N$ is a number of clusters.}
\end{deluxetable}
%*********************************************************************

It is important to note that this model is not applicable to unrelaxed clusters, for which the fundamental 
assumptions in the hydrostatic equilibrium equation do not hold. However, for the full sample, we find that 
the mass bias is nearly constant with very little radial dependence, which is in agreement with previous 
work \citep[e.g.][]{Battaglia:2011a}.  Applying a constant mass correction of 10\% (equivalent to 
using our model with parameter values $\alpha = 0.1$ and $n = 0$), as has been done previously
\citep[e.g.][]{Mantz:2008}, will remove the bias on average for a sample including both relaxed and unrelaxed clusters.  
The scatter in this correction is large, however, at 15--20\%.  Therefore, our correction, though only applicable to a subsample, can provide significant improvements to the cluster mass measurements when such a subsample is applicable, such as gas fraction measurements \citep{Allen:2008,LaRoque:2006}. We note that the correction should in principle be a function of redshift, as the contribution of non-thermal pressure grows to higher redshift when clusters are dynamically younger \citep[e.g.][]{Battaglia:2011a,Shaw:2010}. However, since we do not sample the full accretion histories of clusters at all redshifts, we report best fit values for our model only at $z = 0$.

\section{Discussion and Conclusions}
\label{sec:conclusions}

Using high-resolution adaptive mesh refinement cosmological
simulations of a sample of sixteen clusters, we have performed an
analysis of the effect of major mergers on the hydrostatic mass estimate.  
Rather than examining the properties of galaxy clusters at a single epoch, 
we utilize merger trees to follow the evolution of individual objects during and after mergers. 
We focused specifically on the characteristic features of the hydrostatic mass bias during the
merger and the subsequent evolution of the random ICM motions produced.  

Our principal results can be summarized as follows:
\begin{itemize}

\item Major mergers are the dominant sources of bulk motions providing non-thermal pressure support in the ICM and therefore bias in the hydrostatic mass estimate. 

\item Well after a major merger, the hydrostatic mass estimate underestimates the true mass within $\rc$ by 6--10\%.

\item During a major merger, random kinetic motions contribute $\sim$30\% of the total ICM pressure at $\rc$. Subsequently, the energy in these motions are converted to thermal energy over time, comprising a residual 10\% contribution when the cluster is well relaxed.

\item We provide a method for correcting the hydrostatic mass estimate for relaxed clusters based on the thermal pressure profile that recovers 
the true mass within $\sim$10\% from 0.7--2$\rc$.
\end{itemize}

In detail, we find the evolution of the hydrostatic mass bias can be divided into two epochs. 
During the merger,  the infall of the smaller structure generates asymmetrical shocks that emanate 
from the center of the parent cluster. These shocks result in a mass bias peak that propagates outward 
over the course of a Gyr until it reaches the edge of the cluster.

As the clusters relax, the hydrostatic mass bias  evolves more slowly, decreasing from $\sim$20\% to 
6--10\% within $\rc$.  Even after 8 Gyr, the mass bias in the innermost regions ($<\tfrc$) fluctuates 
around 3\%. The mass bias increases with increasing radius, reaching 18-20\% at $\trc$, which is 
in agreement with previous simulations \citep[e.g.,][]{Kay:2004, Rasia:2006, Nagai:2007a, Lau:2009} 
and observations \citep[e.g.,][]{Mahdavi:2008,Zhang:2010}.

The evolution of the hydrostatic mass bias in the second epoch, 
after the merger, is driven by non-thermal pressure support from bulk motions.  
The random kinetic energy decays after the merger, thermalizing into the surrounding ICM. 
Despited continued accretion, the fraction of total ICM energy in residual bulk motions
falls from $\sim$10\% immediately following the merger to $\sim$5\% long after.  Partly this
reflects the inability of smoothly accreted material to excite new motions in the ICM
and partly the thermalization of existing random motions.
The increase in the thermal energy drives the observed decrease in the hydrostatic mass bias. 

Since the bulk motions in the ICM drive the hydrostatic mass bias, we investigated methods 
for improving the mass estimate by accounting for the non-thermal pressure provided by 
random gas motions. 

Within 4 Gyr of a major merger, a cluster has not yet virialized and is not in equilibrium.  
However,  after this time, a cluster has sufficiently relaxed that with accurate treatment of 
the random pressure, we are able to recover the true mass. We therefore use this criteria 
to split our sample into ``relaxed" and ``unrelaxed" by dividing into clusters that have and 
have not experienced a major merger in the last 4 Gyr.

We provide a method for correcting the hydrostatic mass estimate for relaxed clusters by harnessing 
the similarity in the evolution of the non-thermal pressure profiles. This method combines the true 
thermal pressure profile for each cluster individually with the average non-thermal pressure profile 
of the relaxed sample. At $z\ =\ 0$, our model recovers the true mass to within $\sim$10\% in the 
radial range 0.7--2 $\rc$.

There are a few issues to bear in mind when interpreting our results. 
First, our simulations neglect radiative cooling, energy feedback from stars and active galactic nuclei, 
and additional sources of non-thermal pressure such as cosmic-rays and magnetic field. While
these effects are critical for cluster cores, their effects on cluster outskirts remain uncertain and must be explored in the future studies. Uncertainty in our results may also arise from the lack of viscosity and our treatment of the ICM as a single fluid, neglecting various effects important for plasmas. 
Finally, our measurements use full three dimensional information from the simulations averaged using 
spherical symmetry and therefore do not account for projection effects. We exclude gas contained within 
subhalos of mass greater than $10^{12}\  h^{-1}\ M_{\odot}$, which will differ somewhat from observationally 
identified peaks in emission. These differences may account for our somewhat smaller mass biases than those 
measured from mock X-ray analyses \citep[e.g.][]{Nagai:2007a,Rasia:2004}.  However, despite these caveats, 
we are confident that it is possible to gain better control on the bias in the hydrostatic mass estimate of clusters 
by taking into account its dependence on dynamical state. 
In the future work, we will explore possible morphological classification schemes for defining these 
subsamples based on observational data. 

\acknowledgments 
We thank Erwin Lau and Andrey Kravtsov for useful discussions.  We also acknowledge the anonymous referee for valuable feedback on the manuscript. This work was supported in part by NSF grant AST-1009811, by NASA ATP grant NNX11AE07G, and by the facilities and staff of the Yale University Faculty of Arts and Sciences High Performance Computing Center. D.H.R. acknowledges the support of NSF grant OCI-0904484.  The simulations used in this study were performed on the BulldogM cluster at the Yale University Faculty of Arts and Sciences High Performance Computing Center. 

\bibliographystyle{apj}

\bibliography{ref}

\begin{thebibliography}{42}
\expandafter\ifx\csname natexlab\endcsname\relax\def\natexlab#1{#1}\fi

\bibitem[{Allen {et~al.}(2008)Allen, Rapetti, Schmidt, Ebeling, Morris, \&
  Fabian}]{Allen:2008}
Allen, S.~W., Rapetti, D.~A., Schmidt, R.~W., Ebeling, H., Morris, R.~G., \&
  Fabian, A.~C. 2008, Monthly Notices of the Royal Astronomical Society, 383,
  879

\bibitem[{Battaglia {et~al.}(2011)Battaglia, Bond, Pfrommer, \&
  Sievers}]{Battaglia:2011a}
Battaglia, N., Bond, J.~R., Pfrommer, C., \& Sievers, J.~L. 2011, arXiv,
  astro-ph.CO

\bibitem[{Binney \& Tremaine(2008)}]{Binney:2008}
Binney, J., \& Tremaine, S. 2008, Galactic Dynamics: Second Edition, by James
  Binney and Scott Tremaine. ISBN 978-0-691-13026-2 (HB). Published by
  Princeton University Press, Princeton, NJ USA, 2008., -1

\bibitem[{Gottlober {et~al.}(2001)Gottlober, Klypin, \&
  Kravtsov}]{Gottlober:2001}
Gottlober, S., Klypin, A., \& Kravtsov, A.~V. 2001, \apj, 546, 223

\bibitem[{Iapichino \& Niemeyer(2008)}]{Iapichino:2008}
Iapichino, L., \& Niemeyer, J.~C. 2008, Monthly Notices of the Royal
  Astronomical Society, 388, 1089

\bibitem[{Jeltema {et~al.}(2008)Jeltema, Hallman, Burns, \&
  Motl}]{Jeltema:2008}
Jeltema, T.~E., Hallman, E.~J., Burns, J.~O., \& Motl, P.~M. 2008, arXiv, 681,
  167

\bibitem[{Kaiser(1986)}]{Kaiser:1986}
Kaiser, N. 1986, \mnras, 222, 323

\bibitem[{Kay {et~al.}(2004)Kay, Thomas, Jenkins, \& Pearce}]{Kay:2004}
Kay, S.~T., Thomas, P.~A., Jenkins, A., \& Pearce, F.~R. 2004, \mnras, 355,
  1091

\bibitem[{Klypin {et~al.}(1999)Klypin, Gottlober, Kravtsov, \&
  Khokhlov}]{Klypin:1999}
Klypin, A., Gottlober, S., Kravtsov, A.~V., \& Khokhlov, A.~M. 1999, \apj, 516,
  530

\bibitem[{Klypin {et~al.}(2001)Klypin, Kravtsov, Bullock, \&
  Primack}]{Klypin:2001}
Klypin, A., Kravtsov, A.~V., Bullock, J.~S., \& Primack, J.~R. 2001, \apj, 554,
  903

\bibitem[{Kravtsov(1999)}]{Kravtsov:1999}
Kravtsov, A.~V. 1999, PhD thesis, NEW MEXICO STATE UNIVERSITY

\bibitem[{Kravtsov {et~al.}(2004)Kravtsov, Berlind, Wechsler, Klypin,
  Gottlober, Allgood, \& Primack}]{Kravtsov:2004a}
Kravtsov, A.~V., Berlind, A.~A., Wechsler, R.~H., Klypin, A.~A., Gottlober, S.,
  Allgood, B., \& Primack, J.~R. 2004, \apj, 609, 35

\bibitem[{Kravtsov {et~al.}(2002)Kravtsov, Klypin, \& Hoffman}]{Kravtsov:2002}
Kravtsov, A.~V., Klypin, A., \& Hoffman, Y. 2002, \apj, 571, 563

\bibitem[{LaRoque {et~al.}(2006)LaRoque, Bonamente, Carlstrom, Joy, Nagai,
  Reese, \& Dawson}]{LaRoque:2006}
LaRoque, S.~J., Bonamente, M., Carlstrom, J.~E., Joy, M.~K., Nagai, D., Reese,
  E.~D., \& Dawson, K.~S. 2006, \apj, 652, 917

\bibitem[{Lau {et~al.}(2009)Lau, Kravtsov, \& Nagai}]{Lau:2009}
Lau, E.~T., Kravtsov, A.~V., \& Nagai, D. 2009, \apj, 705, 1129

\bibitem[{Mahdavi {et~al.}(2008)Mahdavi, Hoekstra, Babul, \&
  Henry}]{Mahdavi:2008}
Mahdavi, A., Hoekstra, H., Babul, A., \& Henry, J.~P. 2008, Monthly Notices of
  the Royal Astronomical Society, 384, 1567

\bibitem[{Mantz {et~al.}(2008)Mantz, Allen, Ebeling, \& Rapetti}]{Mantz:2008}
Mantz, A., Allen, S.~W., Ebeling, H., \& Rapetti, D. 2008, Monthly Notices of
  the Royal Astronomical Society, 387, 1179

\bibitem[{Mantz {et~al.}(2010)Mantz, Allen, Rapetti, \& Ebeling}]{Mantz:2010}
Mantz, A., Allen, S.~W., Rapetti, D., \& Ebeling, H. 2010, Monthly Notices of
  the Royal Astronomical Society, 406, no

\bibitem[{Meneghetti {et~al.}(2010)Meneghetti, Rasia, Merten, Bellagamba,
  Ettori, Mazzotta, Dolag, \& Marri}]{Meneghetti:2010}
Meneghetti, M., Rasia, E., Merten, J., Bellagamba, F., Ettori, S., Mazzotta,
  P., Dolag, K., \& Marri, S. 2010, A{\&}A, 514, 93

\bibitem[{Nagai(2006)}]{Nagai:2006}
Nagai, D. 2006, \apj, 650, 538

\bibitem[{Nagai {et~al.}(2007{\natexlab{a}})Nagai, Kravtsov, \&
  Vikhlinin}]{Nagai:2007b}
Nagai, D., Kravtsov, A.~V., \& Vikhlinin, A. 2007{\natexlab{a}}, \apj, 668, 1

\bibitem[{Nagai {et~al.}(2007{\natexlab{b}})Nagai, Vikhlinin, \&
  Kravtsov}]{Nagai:2007a}
Nagai, D., Vikhlinin, A., \& Kravtsov, A.~V. 2007{\natexlab{b}}, \apj, 655, 98

\bibitem[{Piffaretti \& Valdarnini(2008)}]{Piffaretti:2008}
Piffaretti, R., \& Valdarnini, R. 2008, A{\&}A, 491, 71

\bibitem[{Poole {et~al.}(2006)Poole, Fardal, Babul, McCarthy, Quinn, \&
  Wadsley}]{Poole:2006}
Poole, G.~B., Fardal, M.~A., Babul, A., McCarthy, I.~G., Quinn, T., \& Wadsley,
  J. 2006, \mnras, 373, 881

\bibitem[{Puchwein \& Bartelmann(2007)}]{Puchwein:2007}
Puchwein, E., \& Bartelmann, M. 2007, A{\&}A, 474, 745

\bibitem[{Rasia {et~al.}(2011)Rasia, Mazzotta, Evrard, Markevitch, Dolag, \&
  Meneghetti}]{Rasia:2011}
Rasia, E., Mazzotta, P., Evrard, A., Markevitch, M., Dolag, K., \& Meneghetti,
  M. 2011, \apj, 729, 45

\bibitem[{Rasia {et~al.}(2004)Rasia, Tormen, \& Moscardini}]{Rasia:2004}
Rasia, E., Tormen, G., \& Moscardini, L. 2004, \mnras, 351, 237

\bibitem[{Rasia {et~al.}(2006)Rasia, Ettori, Moscardini, Mazzotta, Borgani,
  Dolag, Tormen, Cheng, \& Diaferio}]{Rasia:2006}
Rasia, E., {et~al.} 2006, \mnras, 369, 2013

\bibitem[{Rasia {et~al.}(2012)Rasia, Meneghetti, Martino, Borgani, Bonafede,
  Dolag, Ettori, Fabjan, Giocoli, Mazzotta, Merten, Radovich, \&
  Tornatore}]{Rasia:2012}
---. 2012, eprint arXiv:1201.1569

\bibitem[{Ricker \& Sarazin(2001)}]{Ricker:2001}
Ricker, P.~M., \& Sarazin, C.~L. 2001, \apj, 561, 621

\bibitem[{Ritchie \& Thomas(2002)}]{Ritchie:2002}
Ritchie, B.~W., \& Thomas, P.~A. 2002, Monthly Notices of the Royal
  Astronomical Society, 329, 675

\bibitem[{Rowley {et~al.}(2004)Rowley, Thomas, \& Kay}]{Rowley:2004}
Rowley, D.~R., Thomas, P.~A., \& Kay, S.~T. 2004, Monthly Notices of the Royal
  Astronomical Society, 352, 508

\bibitem[{Rozo {et~al.}(2009)Rozo, Wechsler, Rykoff, Annis, Becker, Evrard,
  Frieman, Hansen, Hao, Johnston, Koester, McKay, Sheldon, \&
  Weinberg}]{Rozo:2010}
Rozo, E., {et~al.} 2009, \apj, 708, 645

\bibitem[{Rudd {et~al.}(2008)Rudd, Zentner, \& Kravtsov}]{Rudd:2008}
Rudd, D.~H., Zentner, A.~R., \& Kravtsov, A.~V. 2008, \apj, 672, 19

\bibitem[{Sehgal {et~al.}(2010)Sehgal, Bode, Das, Hern{\'a}ndez-Monteagudo,
  Huffenberger, Lin, Ostriker, \& Trac}]{Sehgal:2010}
Sehgal, N., Bode, P., Das, S., Hern{\'a}ndez-Monteagudo, C., Huffenberger, K.,
  Lin, Y.-T., Ostriker, J.~P., \& Trac, H. 2010, \apj, 709, 920

\bibitem[{Shaw {et~al.}(2008)Shaw, Holder, \& Bode}]{Shaw:2008}
Shaw, L.~D., Holder, G.~P., \& Bode, P. 2008, \apj, 686, 206

\bibitem[{Shaw {et~al.}(2010)Shaw, Nagai, Bhattacharya, \& Lau}]{Shaw:2010}
Shaw, L.~D., Nagai, D., Bhattacharya, S., \& Lau, E.~T. 2010, arXiv,
  astro-ph.CO

\bibitem[{Vanderlinde {et~al.}(2010)Vanderlinde, Crawford, de~Haan, Dudley,
  Shaw, Ade, Aird, Benson, Bleem, Brodwin, Carlstrom, Chang, Crites, Desai,
  Dobbs, Foley, George, Gladders, Hall, Halverson, High, Holder, Holzapfel,
  Hrubes, Joy, Keisler, Knox, Lee, Leitch, Loehr, Lueker, Marrone, McMahon,
  Mehl, Meyer, Mohr, Montroy, Ngeow, Padin, Plagge, Pryke, Reichardt, Rest,
  Ruel, Ruhl, Schaffer, Shirokoff, Song, Spieler, Stalder, Staniszewski, Stark,
  Stubbs, van Engelen, Vieira, Williamson, Yang, Zahn, \&
  Zenteno}]{Vanderlinde:2010}
Vanderlinde, K., {et~al.} 2010, \apj, 722, 1180

\bibitem[{Vazza {et~al.}(2011)Vazza, Brunetti, Gheller, Brunino, \&
  Br{\"u}ggen}]{Vazza:2011}
Vazza, F., Brunetti, G., Gheller, C., Brunino, R., \& Br{\"u}ggen, M. 2011,
  A{\&}A, 529, A17

\bibitem[{Vikhlinin {et~al.}(2009)Vikhlinin, Burenin, Ebeling, Forman,
  Hornstrup, Jones, Kravtsov, Murray, Nagai, Quintana, \&
  Voevodkin}]{Vikhlinin:2009}
Vikhlinin, A., {et~al.} 2009, \apjs, 692, 1033

\bibitem[{Yang {et~al.}(2010)Yang, Bhattacharya, \& Ricker}]{Yang:2010}
Yang, H.-Y.~K., Bhattacharya, S., \& Ricker, P.~M. 2010, ArXiv e-prints

\bibitem[{Zhang {et~al.}(2010)Zhang, Okabe, Finoguenov, Smith, Piffaretti,
  Valdarnini, Babul, Evrard, Mazzotta, Sanderson, \& Marrone}]{Zhang:2010}
Zhang, Y.-Y., {et~al.} 2010, \apj, 711, 1033

\end{thebibliography}

\end{document}